\documentclass[%
 reprint,
 amsmath,amssymb,
 aps,
]{revtex4-2}

\usepackage{graphicx}
\usepackage{dcolumn}
\usepackage{bm}

\usepackage{physics}
\usepackage{amsmath}
\usepackage{caption}
\usepackage{subcaption}
\begin{document}

\preprint{APS/123-QED}

\title{Nonlinear valley thermal physics in two dimensional materials}

\author{Shivam Sharma}
\author{Abir De Sarkar}%
\altaffiliation{Corresponding author}
\email{abir@inst.ac.in, abirdesarkar@gmail.com}
\affiliation{%
 Institute of Nano Science and Technology, Knowledge City, Sector 81, Mohali, Punjab, 140306, India 
}%

\date{\today}

\begin{abstract}
This study delves into the intrinsic nonlinear valley thermal effects, driven by the system's Quantum Metric. Our findings elucidate that valley indices in the nonlinear effect are distinguishable by the thermoelectric correction to the orbital magnetization, which adopts opposite signs across valleys mirroring the role of orbital angular momentum in the linear valley Hall effect. Through a prototypical two-band model on an anisotropic tilted Dirac semimetal, we investigate how intrinsic material parameters modulate this nonlinear valley thermal response. Extending to realistic $\mathcal{PT}$-symmetric anisotropic semiconductors, our findings enrich the understanding of valley-based phenomena, with implications for advanced theoretical and experimental pursuits in valleytronics and valley-caloritonics.
\end{abstract}

\maketitle



\textit{Introduction.} Valleytronics, akin to spintronics, leverages the valley degree of freedom of electrons for information processing \cite{Laird2013,Rohling2012,Li2013}. By employing external perturbations—magnetic, electrical, thermal, or optical, the information encoded in the valley degree of freedom can be manipulated \cite{Mak2014, Yao2008, Cao2012, MacNeill2015, Li2014}. Of the above-mentioned external perturbations, electric and thermal control is the most promising for practical applications. Thermal control and manipulation of the valley functionality is the pioneer study in research and opens up exciting avenues for its utilization in the realms of valley-caloritronics. Valley-caloritronics presents significant opportunities for information transfer and processing, utilizing the temperature gradient and heat flow omnipresent in all systems. 

The pioneering phenomenon for thermal control of the valley is the linear valley Nernst effect (LVNE), which is a thermal derivative of the linear valley Hall effect (LVHE). In $\mathcal{PT}$-asymmetric systems, an applied electric field induces the separation of electrons with different valley indices, accumulating them at opposite edges of the material, this phenomenon is known as the LVHE \cite{Sokolewicz2019,Xu2021,Wang2024,Guo2024,Dong2023}. Whereas in the case of the LVNE, an applied temperature gradient induces the separation of electrons with different valley indices. The intrinsic LVHE and LVNE arise from the Berry curvature (BC), which is linked to the material's relativistic band structure \cite{Xiao2006,Xiao2012,Yamamoto2015,Yu2015,Dau2019,Du2022}. Notably for $\mathcal{PT}$-asymmetric systems, the BC exhibits equal magnitudes but opposite signs at the two distinct valleys, assigning the opposite transverse velocity to the electrons at the distinct valleys. 

A recent perspective reinterprets the LVNE through the lens of the orbital angular momentum (OAM) of electrons \cite{Bhowal2021,Cysne2021}. In systems lacking inversion symmetry, it has been suggested that the orbital and valley degrees of freedom are intertwined, leading to the accumulation of electrons with opposite OAM at the material's edges \cite{Go2018,Kontani2009,Sharma12024}. The re-interpretation of the LVNE in terms of orbital degree of freedom fails in systems where inversion and time-reversal symmetries are either both broken or both preserved. In $\mathcal{PT}$-symmetric systems, both the BC and intratomic OAM vanishes at every point in the reciprocal space, as both the BC and OAM are odd under $\mathcal{P}$ and even under $\mathcal{T}$ symmetries. This symmetry-imposed constraint suppresses the intrinsic LVNE in $\mathcal{PT}$-symmetric materials that are otherwise highly regarded for their potential in spintronic applications. As a result, the range of materials suitable for valley-based information processing is limited.

The constraint imposed by $\mathcal{PT}$-symmetry may be addressed by investigating second-order nonlinear dynamics of the valley degree of freedom under a thermal gradient. While the intrinsic LVNE has been extensively studied, the nonlinear valley-based transport mechanisms remain less explored. Recently, Das \textit{et al.} demonstrated the intrinsic nonlinear valley Hall effect (NVHE) in tilted massive Dirac fermions \cite{Das2024}. Akin to the nonlinear anomalous Nernst effect, the intrinsic contribution to nonlinear valley Nernst effect (NVNE) arises from the thermoelectric correction to the BC, with the valley-contrasting thermoelectric corrections to the OAM being attributed as the distinction between the two valleys. These corrections, stemming from the thermoelectric response, can be understood in terms of the system’s quantum geometric tensors, which are governed by the momentum-resolved Berry phase of the material \cite{Gao2014,Liu2021}. Notably, the intrinsic NVNE does not require strong spin-orbit coupling, allowing materials typically not considered suitable for spintronics to exhibit a significant NVNE.   

In this letter, we aim to highlight the dissipationless intrinsic NVNE in a $\mathcal{PT}$-symmetric two-dimensional materials. To explore the physics of the intrinsic NVNE, we first examine a prototypical continuum two-band model for type-II Dirac semimetal, deriving an analytical form for the band-resolved quantum metric (QM), a quantum geometric property of electronic wave function, to determine the thermoelectric corrections to the OAM and BC. We also investigate how these corrections depend on the material's intrinsic properties. Beyond the prototypical model, we extend our analysis to realistic two-dimensional materials, such as the $\mathcal{PT}$-symmetric anisotropic semiconductor bilayer graphene with trigonal warping and organic semiconductor $\alpha(BEDT-TTF)_2I_3$, comparing the results with those obtained from the analytical continuum model.

\textit{QM induced intrinsic NVNE}. In quantum kinetic theory, the second-order nonlinear response of the valley current along a specific direction of an applied thermoelectric response is described by \cite{Xiao2021,Xiao12021},
\begin{equation}
    j = -eTr[\hat{\textbf{v}}\rho] + \sum_{q}\textbf{E}^T\times \textbf{M}_q^{\Omega}
\end{equation}
where Tr denotes the trace of the matrix, $\hat{\textbf{v}}$ is the velocity operator, temperature gradient is represented by the thermal field $\textbf{E}^T = -\grad T/T$ and $\textbf{M}_q^{\Omega}$ denotes the Berry curvature induced orbital magnetization density if the $q$-th band \cite{Tatara2015,Xiao2010}. The total Hamiltonian for the system in the presence of thermal gradient is given by, $\mathcal{H} = \mathcal{H}^o + \mathcal{H}^T + \mathcal{H}^D$, here, $\mathcal{H}^T$ and $\mathcal{H}^D$, denotes the temperature gradient and disorder contribution to the unperturbed Hamiltonian $\mathcal{H}^o$, respectively. According to quantum kinetic theory the $N$-th order density matrix in crystal momentum representation $\rho(\textbf{k},t)$, is computed for the Hamiltonian $\mathcal{H}$ through the quantum Liouville equation along with relaxation time approximation \cite{Sekine2017,Culcer2012,Culcer2017,Glazov2014},
\begin{equation}
    \frac{\partial\rho(\textbf{k},t)}{\partial t} + \frac{i}{\hbar}[\mathcal{H}^o,\rho(\textbf{k},t)] + \frac{\rho(\textbf{k},t)}{\tau_s} = D^T(\rho(\textbf{k},t))
\end{equation}
here, $\tau_s$ captures the scattering time of the Bloch electrons and $D^T(\rho)$ denotes the thermal driving term expressed as \cite{Sekine2017}, 
\begin{equation}
    D^T(\rho) = \frac{-1}{2\hbar}\textbf{E}^T.\left(\{\mathcal{H}^o,\partial_{\textbf{k}}\rho\} - i[\mathcal{R}_{\textbf{k}},\{\mathcal{H}^o,\rho\}]\right)
\end{equation}
here, $\{.,.\}$ and $[.,.]$ represents the anticommutaor and commutator brackets, respectively and $\mathcal{R}_k$ is the k-resolved Berry connection \cite{Watanabe2021}. 

Using Eq. (2) the iterative equation for the elements of the $N$-th order density matrix is calculated and is expressed as (see supplemental material (SM) for details),
\begin{equation}
    \rho^{(N)}_{qp} = -i\hbar\frac{[D^T(\rho^{N-1})]_{qp}}{\epsilon_{qp} - (i\hbar N/\tau_s)}
\end{equation}
here, $\epsilon_{qp} = (\epsilon_q - \epsilon_p)$. The details of the calculation for the first and second-order density matrix are provided in SM. The first-order density matrix is decomposed in diagonal and off-diagonal parts as $\rho^{(1)} = \rho_d + \rho_o$ and the second-order density matrix is expressed as, $\rho^{(1)} = \rho_{dd} + \rho_{oo} + \rho_{do} + \rho_{od}$. In this decomposition first subscript denotes the diagonal or off-diagonal part of $\rho^{(2)}$ and the second subscript denotes the diagonal or off-diagonal part of $\rho^{(1)}$. The detailed calculation of the different components of the density matrix is provided in the SM.   

The second-order nonlinear current in the presence of the thermoelectric response is expressed as, $j^{(2)}_{\beta} = \alpha_{\beta;\gamma\delta}E^T_{\gamma}E^T_{\delta}$. Using this relation, the $\tau_s$ independent dissipationless second-order nonlinear conductivity tensor is expressed in terms of the QM as \cite{Varshney2023,Varshney12023},
\begin{equation}
\begin{aligned}
\alpha_{\beta;\gamma\delta}^{T} & = \frac{e}{2\hbar T^{2}}\sum_{q\neq p}\int d\textbf{k}\Big[-(\epsilon_{q}+\epsilon_{p})\{\partial_{\beta}\tilde{\mathcal{G}}_{qp}^{\gamma\delta}-4\partial_{\delta}\tilde{\mathcal{G}}_{qp}^{\beta\gamma}\} \\
 & +2\left(2\partial_{\delta}\mathcal{G}_{qp}^{\beta\gamma}+\tilde{\mathcal{G}}_{qp}^{\beta\gamma}(\partial_{\delta}\epsilon_{p}+5\partial_{\delta}\epsilon_{q})\right) \Big]\epsilon_qf_q . 
\end{aligned}
\end{equation}
here, $\Tilde{\mathcal{G}}_{qp}^{\beta\delta}$ is the band-resolved QM often referred to as Berry curvature polarizability (BCP) \cite{Das2023}. This QM can be further simplified in terms of the relativistic Berry connection ($\mathcal{R}_{qp} = i\bra{u_q}\ket{\grad_k u_p}$), where $u_q$ and $u_p$ are Bloch states. The QM is then expressed as $\Tilde{\mathcal{G}}_{qp}^{\beta\gamma} = \sum_{q\neq m}\frac{Re[\mathcal{R}_{mq}^{\beta}\mathcal{R}_{qm}^{\gamma}]}{\epsilon_{mq}-\epsilon_{qm}}$ and $\mathcal{G}_{qp}^{\beta\gamma} = Re[\mathcal{R}_{pq}^{\beta}\mathcal{R}_{qp}^{\gamma}]$. This third rank conductivity is the $\tau$ independent, therefore intrinsic, stems from the QM and will give rise to the nonlinear Nernst effcet.

The symmetry properties of the QM allow for a finite nonlinear valley Hall response, even in systems where both inversion and time-reversal symmetries are either preserved or broken. Specifically, under time-reversal symmetry, the QM transforms as $\mathcal{T}[\mathcal{G}(k)] = \mathcal{G}(-k)$, and under inversion symmetry, it transforms similarly as $\mathcal{P}[\mathcal{G}(k)] = \mathcal{G}(-k)$. Consequently, in the presence of combined $\mathcal{PT}$-symmetry , the quantum metric remains finite, as $\mathcal{PT}[\mathcal{G}(k)] = \mathcal{G}(k)$. This invariance under $\mathcal{PT}$-symmetry ensures the existence of a finite nonlinear valley Hall response in $\mathcal{PT}$-symmetric systems.

\begin{figure}
    \includegraphics[width=1.15\linewidth]{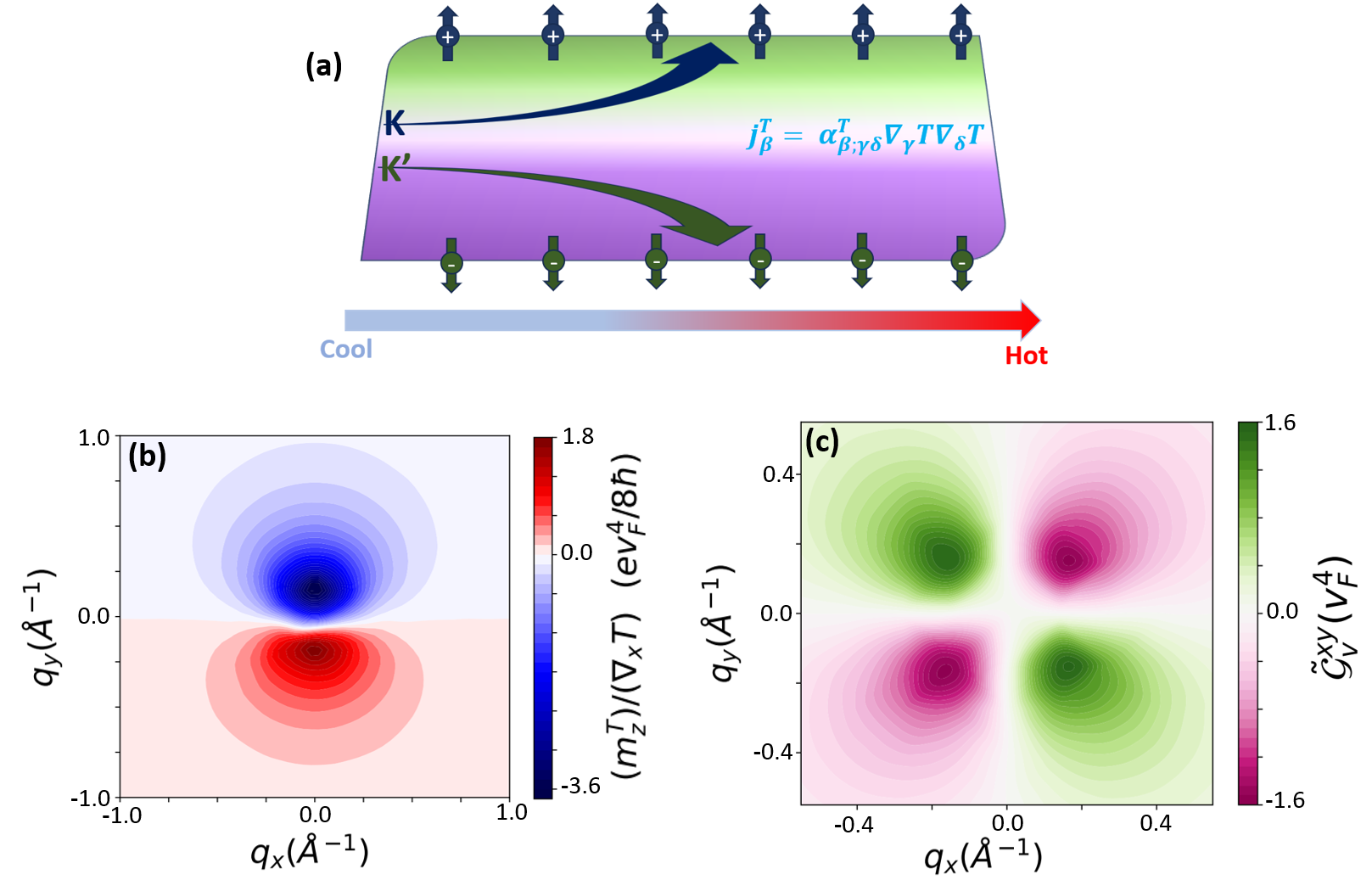}
    \caption{(a) Schematic representation of the nonlinear valley Nernst effect in two-dimensional system induced by the thermal response. The nonlinear valley thermal response if finite in $\mathcal{PT}$-symmetric systems, where the Berry curvature and linear valley Nernst effect vanishes. (b) Depicts the momentum space distribution of the thermal gradient induced OMM at $K$ valley. (c) Depicts the momentum space distribution of the band resolved QM for the valence band.}
    \label{fig:enter-label}
\end{figure}

Intrinsic NVNE is associated with the accumulation of the valley-contrasting temperature gradient induced OAM at the lateral ends of the sample. The temperature gradient induced correction to the orbital magnetic moment (OMM) for the q-th band is expressed as \cite{Xiao2021,Xiao12021},
\begin{equation}
\begin{aligned}
    m_{q,\beta}^{T;o} &= \epsilon_{L}\Big(\sum_{q\neq p}\sum_{q\neq r}e \frac{Re[v_{pr}^\gamma\mathcal{R}^\delta_{rq}\mathcal{R}_{ql}^\eta]}{\epsilon_{qp}}+ \frac{e}{2\hbar}\sum_{q\neq p}(\partial_\gamma \mathcal{G}_{qp}^{\delta\eta}]) \\
     & \quad \quad \quad \quad \quad \quad + \sum_{q\neq p}\frac{1}{2}Re[v_{qq}^\gamma\mathcal{G}_{qp}^{\delta\eta}]\Big)\grad_{\eta}T 
\end{aligned}
\end{equation}
here, the superscript `o' signifies the orbital contribution, $\epsilon_L$ denotes the Levi-Cevita tensor $\epsilon_{\beta\gamma\delta}$. Materials with the valley contrasting temperature gradient induced orbital magnetization ($\mathcal{M}_v^T$) will show a significant intrinsic NVNE.

\textit{Tilted massive Dirac systems.} To analytically study the intrinsic NVNE in $\mathcal{PT}$-symmetric systems, we consider the prototypical continuum model of tilted Dirac semimetal. The low-energy Hamiltonian at the valley indices is expressed as \cite{Hirata2016,Mojarro2021,Mojarro2022};
\begin{equation}
    \mathcal{H}(\textbf{q}) = v_F(\tau\sigma_x q_x + \sigma_yq_y) + \tau \alpha_t q_y\sigma_0 + \Delta \sigma_z
\end{equation}
here, $\textbf{q} = k-\textbf{K}$ defines the momentum relative to the valley point, $v_F$ is the Fermi velocity, and $\sigma$'s represents the Pauli matrices. The parameter $\tau=\pm1$ distinguishes between the two valley indices $K$ and $K^\prime$. The anisotropic term $\alpha_t$ introduces the tilt along the $q_y$ direction. The term $\tau \alpha_t q_y$ breaks both the $\mathcal{P}$ and $\mathcal{T}$ symmetries at the two valleys, and $\Delta$ denotes the band gap of the system. The energy dispersion for the Hamiltonian (7) is given by $\epsilon_{v(c)}=\tau\alpha_t q_y \pm \epsilon$, here $\epsilon = \sqrt{v_F^2q^2 + \Delta^2}$. 

The presence of the $\mathcal{PT}$-symmetry results in the vanishing BC and the suppression of an LVNE in the Hamiltonian (7), nevertheless, an QM-induced nonlinear response can still manifest. The various components of the band resolved QM for the valence band, calculated for the Hamiltonian (4) are,
\begin{equation}
	\begin{pmatrix}
		\tilde{\mathcal{G}}_{vc}^{xx} & \tilde{\mathcal{G}}_{vc}^{xy}\\
		\tilde{\mathcal{G}}_{vc}^{yx} & \tilde{\mathcal{G}}_{vc}^{yy}
	\end{pmatrix} 
 =
 -\frac{v_F^2}{8\epsilon^5}\begin{pmatrix}
     v_F^2q_y^2 + \Delta^2 & -v_F^2q_xq_y\\
     -v_F^2q_xq_y & v_F^2q_x^2 + \Delta^2
 \end{pmatrix}
\end{equation}
The QM for the valence band is notably unaffected by the terms that break either $\mathcal{P}$ or $\mathcal{T}$ symmetry. As a result, the quantum metric can remain finite even when both $\mathcal{P}$ and $\mathcal{T}$ symmetries are preserved, contrary to the behavior of the BC and OAM. This persistence of a finite QM plays a crucial role in driving the NVNE in materials that exhibit combined $\mathcal{PT}$ symmetry. Moreover, the QM's independence from the valley index parameter $\tau$ ensures that it maintains the same sign across both $K$ and $K^\prime$ valleys. 

\begin{figure}
    \includegraphics[width=1.05\linewidth]{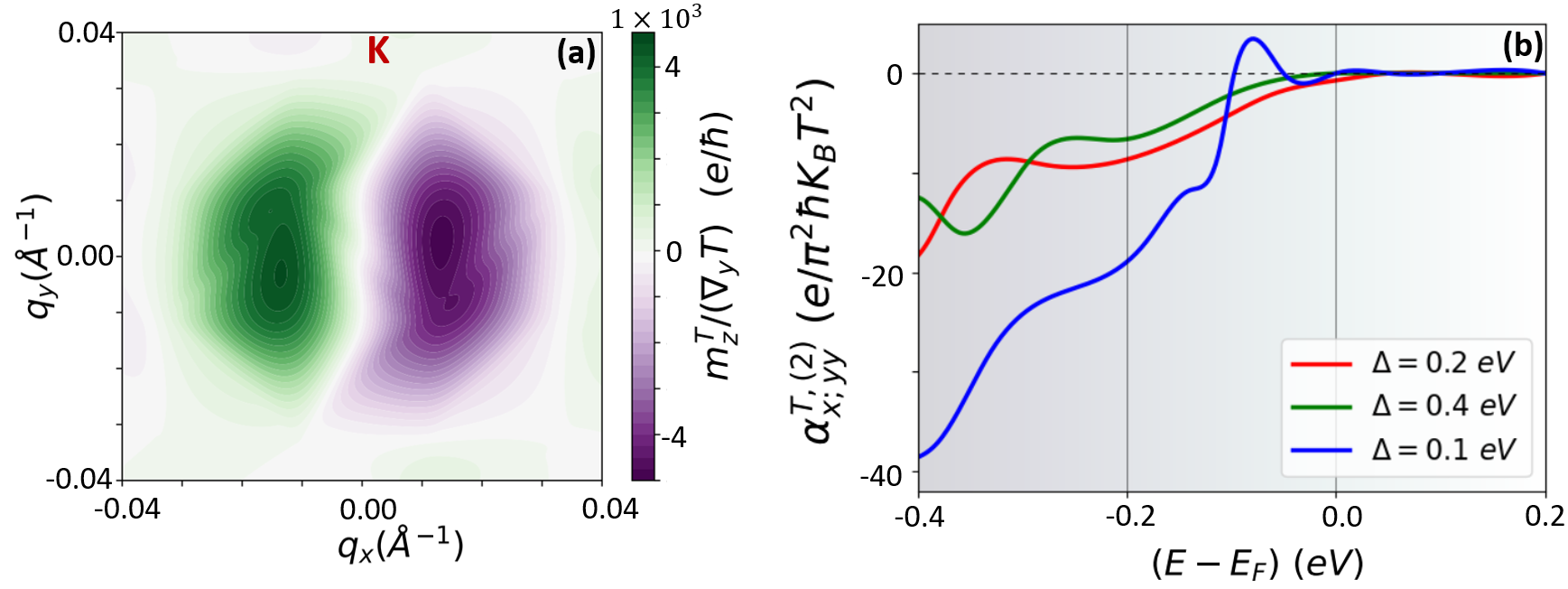}
    \caption{(a) Present the momentum space distribution of the thermal gradient induced orbital magnetic moment (OMM) at the $K$ valley in bilayer graphene (for the calculations we have used, $\Tilde{\Delta} = 0.1 eV, \chi = -0.102 eV \AA^2, \kappa = 0.853 eV \AA^3$, and $\gamma = 0.382 eV \AA$). (b) Depicts the variation in the NVNC as a function of Fermi energy for the different values of band gap energy. Here, the Fermi energy ($E_F$) is set to be 0.1 eV above the VBM.}
    \label{fig:enter-label}
\end{figure}

The QM induced intrinsic valley and band-resolved NVNC with varying chemical potential ($\mu_c$) is calculated using the Eq. (5), for the analytical calculations we have assumed $v_F >> \alpha_t \ \& \ \Delta$ (for detailed calculation refer to the supplemental material (SM)),
\begin{equation}
    \alpha_{y;xx}^{T,(2)}(\zeta) = \zeta\frac{e\tau\alpha_t}{64\pi\hbar T^2}\left[17+29\frac{\Delta^4}{\mu_c^4}-46\frac{\Delta^2}{\mu_c^2}\right]
\end{equation}
here, $\zeta = +1(-1)$ specifies the band index for the conduction (valence) band. The dependence of the NVNC over $\tau$ emphasizes the valley-resolved Nernst response and the accumulation of electrons on the opposite ends based on the valley index. Interestingly, as the $\alpha_t \to 0$, the NVNC vanishes, therefore, the band anisotropy is necessary for the realization of the NVNC in $\mathcal{PT}$-symmetric system. As the band gap term $\Delta \to 0$, the NVNC attains its maximum value, given by $\alpha_{y;xx}^{T,(2);max}(\zeta) = \zeta [17 e\alpha_t/(64\pi\hbar T^2)]$, for the gapless systems the NVNC will be independent of variation in $\mu_c$, whereas for the NVHE, the conductivity is varies inversely as the square of $\mu_c$ \cite{Das2024}. The NLVHC is directly linked to the anisotropy in band dispersion term $\alpha$, the greater the band dispersion anisotropy, the higher the NLVHC. 

The valley-contrasting thermoelectric correction induced OMM for the valence band is calculated using Eq. (6) \& (8) (for detailed calculation, refer to the SM),
\begin{equation}
    \frac{m_{v,z}^{T,o}}{\grad_xT} = -\frac{e v_F^4q_y}{8\hbar\epsilon_o^4}\left[1 + 2\frac{\tau\alpha_t (v_F^2 q_y^2 + \Delta^2)}{\epsilon_o v_F^2q_y}\right]
\end{equation}
The explicit dependence on $\tau$ depicts the valley-contrasting nature of $m_z^{T,o}$. Interestingly, Eq. (10) emphasizes the crucial role of the band dispersion anisotropy in intrinsic NVNE, in the absence of this anisotropy the valley-contrasting $m_z^{T,o}$ disappears and the intrinsic NVNC is nullified. Thus, in addition to $\mathcal{PT}$-symmetry, anisotropy is essential for a significant intrinsic NVNE to emerge. The temperature gradient induced orbital magnetization ($\mathcal{M}^T_v=\int_0^{k_F} m_{v,z}^{T,o} d\textbf{K}$), the values of at $K$ and $K^{\prime}$ is (detailed calculations are incorporated in SM),
\begin{equation}
    \mathcal{M}^{T}_v = \frac{e\tau\alpha_t}{16\pi\hbar}\Big[\frac{1}{\epsilon_F}+3\frac{\Delta^2}{\epsilon_F^3}\Big]\grad_xT
\end{equation} 
The sign of $\mathcal{M}^{T}_v$ is opposite at the two valley points. Interestingly, in the limit $\Delta \to 0$, $\mathcal{M}^{T}_v$ is finite and varies inversely as a function of Fermi energy. 

\textit{Bilayer graphene with trigonal warping.} To demonstrate the robustness of the analytical prototypical antiferromagnetic model, we consider working out with the realistic $\mathcal{PT}$-symmetric material, for this purpose we consider $\mathcal{PT}$-symmetric bilayer graphene. In bilayer graphene, trigonal warping induces a breaking of both $\mathcal{P}$ and $\mathcal{T}$ symmetries simultaneously \cite{Jr2009}. Although the Berry curvature in each graphene layer is equal in magnitude and opposite in sign, resulting in a net zero Berry curvature and vanishing LVNE. However, the anisotropy from trigonal warping can yield nonzero QM's. This intrinsic feature gives rise to a finite NVNE in the bilayer graphene system. 

The two band low energy effective Hamiltonian can be obtained with Lodwing partitioning \cite{Jr2009,Cserti2007,Andor2013,Wu2021,Winkler2003,Kechedzhi2007},
\begin{equation*}
    H^{TW} = H_0 + H_{3w} + H_{cub}
\end{equation*}
\begin{equation}
\begin{aligned}
       H^{TW} & = \begin{pmatrix}
       \Tilde{\Delta} & \tau\gamma q_-\\
       \tau\gamma q_+ & -\Tilde{\Delta} 
   \end{pmatrix} + \chi\begin{pmatrix}
        0 & (q_+)^2 \\
       (q_-)^2 & 0 
   \end{pmatrix} \\
   & \quad -\tau\kappa q^2/2\begin{pmatrix}
       0 & q_-\\
       q_+ & 0
   \end{pmatrix}
\end{aligned}
\end{equation}
where, the parameters $\chi, \kappa$ can be expressed in terms of $\Tilde{\Delta}$ and $\gamma$, the trigonal warping is exhibited by the $\chi$ term and the cubic term is used for cubic fitting of the bands away from the valley points. $q_{\pm} = q_x \pm i\tau q_y$ measures the relative distance from the two valley points. The energy eigenvalue for the $H^{TW}$ are $\epsilon_{v(c)} = \pm\epsilon = \pm\sqrt{\Tilde{\Delta}^2+|g|^2}$, here $g(q) = \tau(\gamma-\kappa q^2/2)q_-+\chi (q_+)^2$. The band gap for the Hamiltonian (12) is given by the $2\Tilde{\Delta}$ term.

The thermal gradient induced OMM at the $K$ valley for the Hamiltonian is depicted in Fig. 2 (a), the overall behavior and magnitude of the OMM are the same at $K$ and $K^{\prime}$. The electric field induced orbital magnetization can be calculated in the same way as discussed for the prototypical two-band model, the calculated OMM attains the value of $\mathcal{M}_z^{T}(K)=-444.512 \times (e \grad_y T/\hbar) \ \& \ \mathcal{M}_z^{T,o}(K^{\prime})=444.512 \times (e \grad_y T/\hbar)$ at $K^{\prime} \ \& \ K$ valleys respectively. The valley-contrasting behavior of the orbital magnetization could be witnessed which is necessary for distinguishing the two valleys. In bilayer graphene, this characteristic facilitates a significant intrinsic NVNC, shown as a function of the Fermi level in Fig. 2 (b). As depicted, NVNE decreases with increasing $\Tilde{\Delta}$, suggesting that small-gap anisotropic materials are ideal candidates for achieving NLVHE in non-magnetic $\mathcal{PT}$-symmetric systems.

$\mathcal{PT}$\textit{-symmetric organic semiconductor.} To demonstrate the robustness of the analytical prototypical continuum model, we consider working out with the realistic $\mathcal{PT}$-symmetric material, for this purpose we have chosen the $\mathcal{PT}$-symmetric anisotropic organic semiconductor $\alpha(BEDT-TTF)_2I_3$ with tilted Dirac cones. The effective tight-binding Hamiltonian can be written in the basis $\{\phi_{A}, \phi_{B}\}$ for two sublattices A and B, and expressed as $\mathcal{H}(K) = [\{h(k) \ \ g^{*}(k)\}, \ \{g(k) \ \ h(k)\}]$ \cite{Goerbig2008,Kobayashi2007}.
Here, $h(k) = 2[t_1e^{i(k_x+k_y)/2} + t_1^{\prime}e^{-i(k_x+k_y)/2} + t_2e^{i(k_x-k_y)/2} + t_2^{\prime}e^{-i(k_x-k_y)/2}$, and $g(k)=t_n(e^{ik_q}+e^{-ik_y})$. $t_i,t_1^{\prime}$ and $t_n$ are the nearest-neighbor and next-nearest-neighbor hopping terms, respectively.

The correction to the OMM induced by a thermal gradient is illustrated in Fig. 3 (c). By evaluating the thermal response induced orbital magnetization at the two valley points and considering a circular region (as depicted with the highlighted circle in Fig. 3(c)) centered at each valley point, we find $\mathcal{M}^T_v(K) = 28.142 \times (e \grad_y/\hbar)$ and $\mathcal{M}^T_v(K^{\prime}) = -28.142 \times (e \grad_y/2\hbar)$, this opposite value $\mathcal{M}_v^T$ at the two valley indices acts as the distinction between two valley points. Fig. 3 (d) shows the variation of the NVNC with the varying Fermi level, as depicted in the plot the NVNC is opposite when the $E_F$ lies in the valence band compared to the conduction band.  

\begin{figure}
    \centering
		\begin{subfigure}{0.5\textwidth}
			\includegraphics[width=\textwidth]{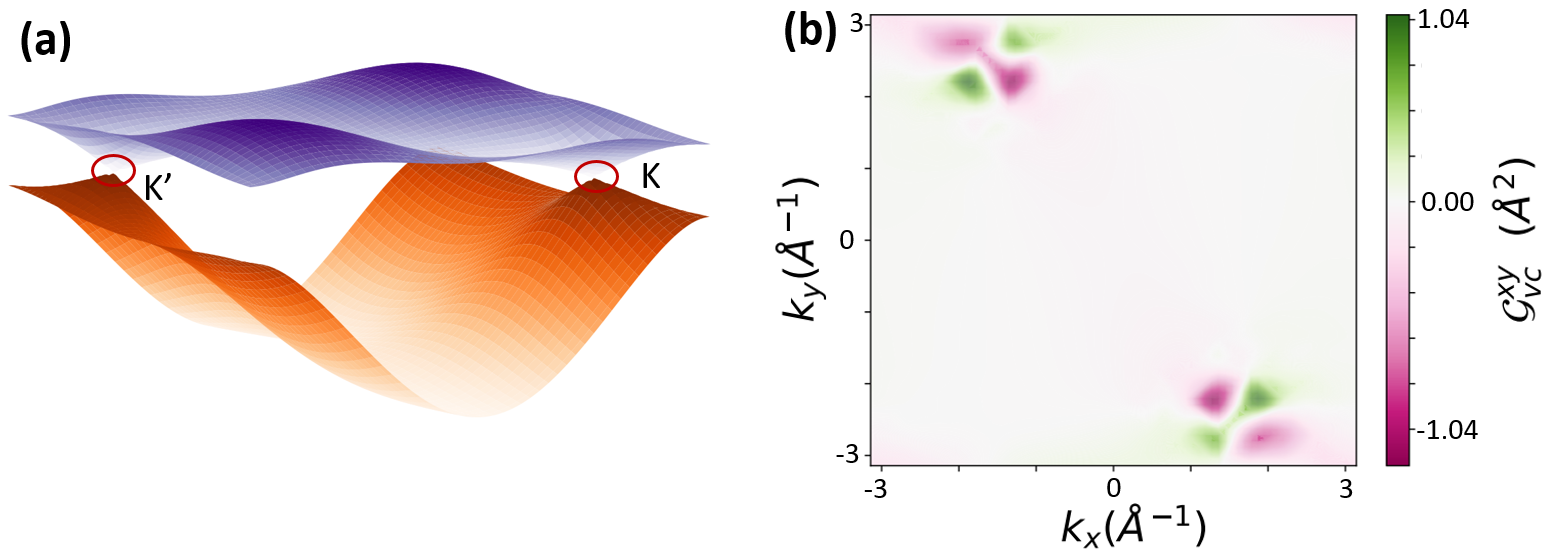}
			\label{fig:first}
		\end{subfigure}
		\hfill
		\begin{subfigure}{0.475\textwidth}
			\includegraphics[width=\textwidth]{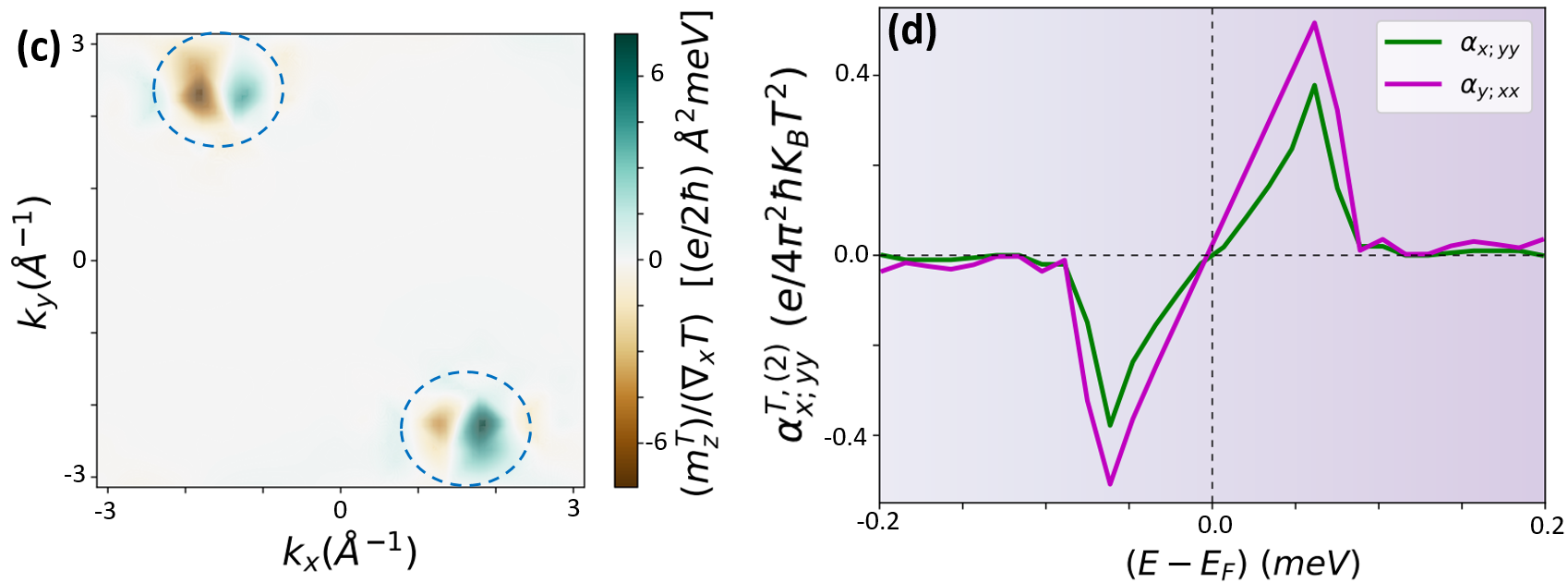}
			\label{fig:second}
		\end{subfigure}
    \caption{(a) The energy dispersion for the anisotrpic organic semiconductor $\alpha(BEDT-TTF)_2I_3$. (b) Depicts the momentum space distribution of the QM ($\mathcal{G}_{vc}^{xy}$). (c) Present the momentum space distribution of the thermal gradient induced orbital magnetic moment (OMM) in $\alpha(BEDT-TTF)_2I_3$. (d) Depicts the variation in the NVNC ($\alpha_{x;yy}^{T}$ and $\alpha_{y;xx}^{T}$) as a function of varying Fermi energy. Calculation are performed assuming the following values of the parameters $\{t_1:36; \ t_1^{\prime}:-86; \ t_2=-24; \ t_2^{\prime}:-77; \ t_n:-60\}$ in meV}.
    \label{fig:enter-label}
\end{figure}

\textit{Conclusion.}
In summary, we have predicted a significant intrinsic nonlinear valley thermal effect in $\mathcal{PT}$-symmetric systems, where the linear valley Nernst and Hall effects are notably absent. Analytical studies, using a prototypical continuum model, reveal that the quantum-metric induced orbital magnetization exhibits a valley-contrasting behavior, providing a means to distinguish between valley indices. Extending beyond this model, our investigation into $\mathcal{PT}$-symmetric bilayer graphene and anisotropic semiconductors $\alpha(BEDT-TTF)_2I_3$ confirms the presence of NVNE in realistic systems. These findings offer a theoretical foundation for manipulating the valley degree of freedom, with potential implications for valleytronics and valley-caloritronics. Furthermore, the NVNE framework may inspire exploration of the extension of nonlinear effects in bosonic systems, opening avenues for investigating magnonic contributions to valley-related phenomena \cite{Xiao2007,Yu2015,Neumann2022,Sharma2024}. Additionally, nontrivial physics is likely to emerge in spin-orbit coupled systems where the spin-valley coupling has been shown to impact both the
spin and valley Nernst effect. 

\pagebreak

\textit{Acknowledgment}

The authors express gratitude to the Institute of Nano Science and Technology (INST), Mohali, India, for providing fellowship support. Support from Project No. CRG/2023/001896 of Science and Engineering Research Board – Department of Science \& Technology (SERB-DST), Govt. of India is gratefully acknowledged.


\nocite{*}

\begin{thebibliography}{53}%
	\makeatletter
	\providecommand \@ifxundefined [1]{%
		\@ifx{#1\undefined}
	}%
	\providecommand \@ifnum [1]{%
		\ifnum #1\expandafter \@firstoftwo
		\else \expandafter \@secondoftwo
		\fi
	}%
	\providecommand \@ifx [1]{%
		\ifx #1\expandafter \@firstoftwo
		\else \expandafter \@secondoftwo
		\fi
	}%
	\providecommand \natexlab [1]{#1}%
	\providecommand \enquote  [1]{``#1''}%
	\providecommand \bibnamefont  [1]{#1}%
	\providecommand \bibfnamefont [1]{#1}%
	\providecommand \citenamefont [1]{#1}%
	\providecommand \href@noop [0]{\@secondoftwo}%
	\providecommand \href [0]{\begingroup \@sanitize@url \@href}%
	\providecommand \@href[1]{\@@startlink{#1}\@@href}%
	\providecommand \@@href[1]{\endgroup#1\@@endlink}%
	\providecommand \@sanitize@url [0]{\catcode `\\12\catcode `\$12\catcode
		`\&12\catcode `\#12\catcode `\^12\catcode `\_12\catcode `\%12\relax}%
	\providecommand \@@startlink[1]{}%
	\providecommand \@@endlink[0]{}%
	\providecommand \url  [0]{\begingroup\@sanitize@url \@url }%
	\providecommand \@url [1]{\endgroup\@href {#1}{\urlprefix }}%
	\providecommand \urlprefix  [0]{URL }%
	\providecommand \Eprint [0]{\href }%
	\providecommand \doibase [0]{https://doi.org/}%
	\providecommand \selectlanguage [0]{\@gobble}%
	\providecommand \bibinfo  [0]{\@secondoftwo}%
	\providecommand \bibfield  [0]{\@secondoftwo}%
	\providecommand \translation [1]{[#1]}%
	\providecommand \BibitemOpen [0]{}%
	\providecommand \bibitemStop [0]{}%
	\providecommand \bibitemNoStop [0]{.\EOS\space}%
	\providecommand \EOS [0]{\spacefactor3000\relax}%
	\providecommand \BibitemShut  [1]{\csname bibitem#1\endcsname}%
	\let\auto@bib@innerbib\@empty
	\bibitem [{\citenamefont {Laird}\ \emph {et~al.}(2013)\citenamefont {Laird},
		\citenamefont {Pei},\ and\ \citenamefont {Kouwenhoven}}]{Laird2013}%
	\BibitemOpen
	\bibfield  {author} {\bibinfo {author} {\bibfnamefont {E.~A.}\ \bibnamefont
			{Laird}}, \bibinfo {author} {\bibfnamefont {F.}~\bibnamefont {Pei}},\ and\
		\bibinfo {author} {\bibfnamefont {L.~P.}\ \bibnamefont {Kouwenhoven}},\
	}\bibfield  {title} {\bibinfo {title} {A valley–spin qubit in a carbon
			nanotube},\ }\href {https://doi.org/10.1038/nnano.2013.140} {\bibfield
		{journal} {\bibinfo  {journal} {Nature Nanotechnology}\ }\textbf {\bibinfo
			{volume} {8}},\ \bibinfo {pages} {565} (\bibinfo {year} {2013})}\BibitemShut
	{NoStop}%
	\bibitem [{\citenamefont {Rohling}\ and\ \citenamefont
		{Burkard}(2012)}]{Rohling2012}%
	\BibitemOpen
	\bibfield  {author} {\bibinfo {author} {\bibfnamefont {N.}~\bibnamefont
			{Rohling}}\ and\ \bibinfo {author} {\bibfnamefont {G.}~\bibnamefont
			{Burkard}},\ }\bibfield  {title} {\bibinfo {title} {Universal quantum
			computing with spin and valley states},\ }\href
	{https://doi.org/10.1088/1367-2630/14/8/083008} {\bibfield  {journal}
		{\bibinfo  {journal} {New Journal of Physics}\ }\textbf {\bibinfo {volume}
			{14}},\ \bibinfo {pages} {83008} (\bibinfo {year} {2012})}\BibitemShut
	{NoStop}%
	\bibitem [{\citenamefont {Li}\ \emph {et~al.}(2013)\citenamefont {Li},
		\citenamefont {Cao}, \citenamefont {Niu}, \citenamefont {Shi},\ and\
		\citenamefont {Feng}}]{Li2013}%
	\BibitemOpen
	\bibfield  {author} {\bibinfo {author} {\bibfnamefont {X.}~\bibnamefont
			{Li}}, \bibinfo {author} {\bibfnamefont {T.}~\bibnamefont {Cao}}, \bibinfo
		{author} {\bibfnamefont {Q.}~\bibnamefont {Niu}}, \bibinfo {author}
		{\bibfnamefont {J.}~\bibnamefont {Shi}},\ and\ \bibinfo {author}
		{\bibfnamefont {J.}~\bibnamefont {Feng}},\ }\bibfield  {title} {\bibinfo
		{title} {Coupling the valley degree of freedom to antiferromagnetic order},\
	}\href {https://doi.org/10.1073/pnas.1219420110} {\bibfield  {journal}
		{\bibinfo  {journal} {Proceedings of the National Academy of Sciences}\
		}\textbf {\bibinfo {volume} {110}},\ \bibinfo {pages} {3738} (\bibinfo {year}
		{2013})}\BibitemShut {NoStop}%
	\bibitem [{\citenamefont {Mak}\ \emph {et~al.}(2014)\citenamefont {Mak},
		\citenamefont {McGill}, \citenamefont {Park},\ and\ \citenamefont
		{McEuen}}]{Mak2014}%
	\BibitemOpen
	\bibfield  {author} {\bibinfo {author} {\bibfnamefont {K.~F.}\ \bibnamefont
			{Mak}}, \bibinfo {author} {\bibfnamefont {K.~L.}\ \bibnamefont {McGill}},
		\bibinfo {author} {\bibfnamefont {J.}~\bibnamefont {Park}},\ and\ \bibinfo
		{author} {\bibfnamefont {P.~L.}\ \bibnamefont {McEuen}},\ }\bibfield  {title}
	{\bibinfo {title} {The valley hall effect in $mos_2$ transistors},\ }\href
	{https://doi.org/10.1126/science.1250140} {\bibfield  {journal} {\bibinfo
			{journal} {Science}\ }\textbf {\bibinfo {volume} {344}},\ \bibinfo {pages}
		{1489} (\bibinfo {year} {2014})}\BibitemShut {NoStop}%
	\bibitem [{\citenamefont {Yao}\ \emph {et~al.}(2008)\citenamefont {Yao},
		\citenamefont {Xiao},\ and\ \citenamefont {Niu}}]{Yao2008}%
	\BibitemOpen
	\bibfield  {author} {\bibinfo {author} {\bibfnamefont {W.}~\bibnamefont
			{Yao}}, \bibinfo {author} {\bibfnamefont {D.}~\bibnamefont {Xiao}},\ and\
		\bibinfo {author} {\bibfnamefont {Q.}~\bibnamefont {Niu}},\ }\bibfield
	{title} {\bibinfo {title} {Valley-dependent optoelectronics from inversion
			symmetry breaking},\ }\href {https://doi.org/10.1103/PhysRevB.77.235406}
	{\bibfield  {journal} {\bibinfo  {journal} {Physical Review B}\ }\textbf
		{\bibinfo {volume} {77}},\ \bibinfo {pages} {235406} (\bibinfo {year}
		{2008})}\BibitemShut {NoStop}%
	\bibitem [{\citenamefont {Cao}\ \emph {et~al.}(2012)\citenamefont {Cao},
		\citenamefont {Wang}, \citenamefont {Han}, \citenamefont {Ye}, \citenamefont
		{Zhu}, \citenamefont {Shi}, \citenamefont {Niu}, \citenamefont {Tan},
		\citenamefont {Wang}, \citenamefont {Liu},\ and\ \citenamefont
		{Feng}}]{Cao2012}%
	\BibitemOpen
	\bibfield  {author} {\bibinfo {author} {\bibfnamefont {T.}~\bibnamefont
			{Cao}}, \bibinfo {author} {\bibfnamefont {G.}~\bibnamefont {Wang}}, \bibinfo
		{author} {\bibfnamefont {W.}~\bibnamefont {Han}}, \bibinfo {author}
		{\bibfnamefont {H.}~\bibnamefont {Ye}}, \bibinfo {author} {\bibfnamefont
			{C.}~\bibnamefont {Zhu}}, \bibinfo {author} {\bibfnamefont {J.}~\bibnamefont
			{Shi}}, \bibinfo {author} {\bibfnamefont {Q.}~\bibnamefont {Niu}}, \bibinfo
		{author} {\bibfnamefont {P.}~\bibnamefont {Tan}}, \bibinfo {author}
		{\bibfnamefont {E.}~\bibnamefont {Wang}}, \bibinfo {author} {\bibfnamefont
			{B.}~\bibnamefont {Liu}},\ and\ \bibinfo {author} {\bibfnamefont
			{J.}~\bibnamefont {Feng}},\ }\bibfield  {title} {\bibinfo {title}
		{Valley-selective circular dichroism of monolayer molybdenum disulphide},\
	}\href {https://doi.org/10.1038/ncomms1882} {\bibfield  {journal} {\bibinfo
			{journal} {Nature Communications}\ }\textbf {\bibinfo {volume} {3}},\
		\bibinfo {pages} {887} (\bibinfo {year} {2012})}\BibitemShut {NoStop}%
	\bibitem [{\citenamefont {MacNeill}\ \emph {et~al.}(2015)\citenamefont
		{MacNeill}, \citenamefont {Heikes}, \citenamefont {Mak}, \citenamefont
		{Anderson}, \citenamefont {Kormányos}, \citenamefont {Zólyomi},
		\citenamefont {Park},\ and\ \citenamefont {Ralph}}]{MacNeill2015}%
	\BibitemOpen
	\bibfield  {author} {\bibinfo {author} {\bibfnamefont {D.}~\bibnamefont
			{MacNeill}}, \bibinfo {author} {\bibfnamefont {C.}~\bibnamefont {Heikes}},
		\bibinfo {author} {\bibfnamefont {K.~F.}\ \bibnamefont {Mak}}, \bibinfo
		{author} {\bibfnamefont {Z.}~\bibnamefont {Anderson}}, \bibinfo {author}
		{\bibfnamefont {A.}~\bibnamefont {Kormányos}}, \bibinfo {author}
		{\bibfnamefont {V.}~\bibnamefont {Zólyomi}}, \bibinfo {author}
		{\bibfnamefont {J.}~\bibnamefont {Park}},\ and\ \bibinfo {author}
		{\bibfnamefont {D.~C.}\ \bibnamefont {Ralph}},\ }\bibfield  {title} {\bibinfo
		{title} {Breaking of valley degeneracy by magnetic field in monolayer
			$mose_2$},\ }\href {https://doi.org/10.1103/PhysRevLett.114.037401}
	{\bibfield  {journal} {\bibinfo  {journal} {Physical Review Letters}\
		}\textbf {\bibinfo {volume} {114}},\ \bibinfo {pages} {37401} (\bibinfo
		{year} {2015})}\BibitemShut {NoStop}%
	\bibitem [{\citenamefont {Li}\ \emph {et~al.}(2014)\citenamefont {Li},
		\citenamefont {Ludwig}, \citenamefont {Low}, \citenamefont {Chernikov},
		\citenamefont {Cui}, \citenamefont {Arefe}, \citenamefont {Kim},
		\citenamefont {van~der Zande}, \citenamefont {Rigosi}, \citenamefont {Hill},
		\citenamefont {Kim}, \citenamefont {Hone}, \citenamefont {Li}, \citenamefont
		{Smirnov},\ and\ \citenamefont {Heinz}}]{Li2014}%
	\BibitemOpen
	\bibfield  {author} {\bibinfo {author} {\bibfnamefont {Y.}~\bibnamefont
			{Li}}, \bibinfo {author} {\bibfnamefont {J.}~\bibnamefont {Ludwig}}, \bibinfo
		{author} {\bibfnamefont {T.}~\bibnamefont {Low}}, \bibinfo {author}
		{\bibfnamefont {A.}~\bibnamefont {Chernikov}}, \bibinfo {author}
		{\bibfnamefont {X.}~\bibnamefont {Cui}}, \bibinfo {author} {\bibfnamefont
			{G.}~\bibnamefont {Arefe}}, \bibinfo {author} {\bibfnamefont {Y.~D.}\
			\bibnamefont {Kim}}, \bibinfo {author} {\bibfnamefont {A.~M.}\ \bibnamefont
			{van~der Zande}}, \bibinfo {author} {\bibfnamefont {A.}~\bibnamefont
			{Rigosi}}, \bibinfo {author} {\bibfnamefont {H.~M.}\ \bibnamefont {Hill}},
		\bibinfo {author} {\bibfnamefont {S.~H.}\ \bibnamefont {Kim}}, \bibinfo
		{author} {\bibfnamefont {J.}~\bibnamefont {Hone}}, \bibinfo {author}
		{\bibfnamefont {Z.}~\bibnamefont {Li}}, \bibinfo {author} {\bibfnamefont
			{D.}~\bibnamefont {Smirnov}},\ and\ \bibinfo {author} {\bibfnamefont {T.~F.}\
			\bibnamefont {Heinz}},\ }\bibfield  {title} {\bibinfo {title} {Valley
			splitting and polarization by the zeeman effect in monolayer $mose_2$},\
	}\href {https://doi.org/10.1103/PhysRevLett.113.266804} {\bibfield  {journal}
		{\bibinfo  {journal} {Physical Review Letters}\ }\textbf {\bibinfo {volume}
			{113}},\ \bibinfo {pages} {266804} (\bibinfo {year} {2014})}\BibitemShut
	{NoStop}%
	\bibitem [{\citenamefont {Sokolewicz}\ \emph {et~al.}(2019)\citenamefont
		{Sokolewicz}, \citenamefont {Ghosh}, \citenamefont {Yudin}, \citenamefont
		{Manchon},\ and\ \citenamefont {Titov}}]{Sokolewicz2019}%
	\BibitemOpen
	\bibfield  {author} {\bibinfo {author} {\bibfnamefont {R.}~\bibnamefont
			{Sokolewicz}}, \bibinfo {author} {\bibfnamefont {S.}~\bibnamefont {Ghosh}},
		\bibinfo {author} {\bibfnamefont {D.}~\bibnamefont {Yudin}}, \bibinfo
		{author} {\bibfnamefont {A.}~\bibnamefont {Manchon}},\ and\ \bibinfo {author}
		{\bibfnamefont {M.}~\bibnamefont {Titov}},\ }\bibfield  {title} {\bibinfo
		{title} {Spin-orbit torques in a rashba honeycomb antiferromagnet},\ }\href
	{https://doi.org/10.1103/PhysRevB.100.214403} {\bibfield  {journal} {\bibinfo
			{journal} {Physical Review B}\ }\textbf {\bibinfo {volume} {100}},\ \bibinfo
		{pages} {214403} (\bibinfo {year} {2019})}\BibitemShut {NoStop}%
	\bibitem [{\citenamefont {Xu}\ \emph {et~al.}(2021)\citenamefont {Xu},
		\citenamefont {He}, \citenamefont {Dai}, \citenamefont {Huang},\ and\
		\citenamefont {Ma}}]{Xu2021}%
	\BibitemOpen
	\bibfield  {author} {\bibinfo {author} {\bibfnamefont {X.}~\bibnamefont
			{Xu}}, \bibinfo {author} {\bibfnamefont {Z.}~\bibnamefont {He}}, \bibinfo
		{author} {\bibfnamefont {Y.}~\bibnamefont {Dai}}, \bibinfo {author}
		{\bibfnamefont {B.}~\bibnamefont {Huang}},\ and\ \bibinfo {author}
		{\bibfnamefont {Y.}~\bibnamefont {Ma}},\ }\bibfield  {title} {\bibinfo
		{title} {Single-valley state in a two-dimensional antiferromagnetic
			lattice},\ }\href {https://doi.org/10.1103/PhysRevB.104.205430} {\bibfield
		{journal} {\bibinfo  {journal} {Physical Review B}\ }\textbf {\bibinfo
			{volume} {104}},\ \bibinfo {pages} {205430} (\bibinfo {year}
		{2021})}\BibitemShut {NoStop}%
	\bibitem [{\citenamefont {Wang}\ \emph {et~al.}(2024)\citenamefont {Wang},
		\citenamefont {Feng}, \citenamefont {Dai}, \citenamefont {Huang},
		\citenamefont {Ma},\ and\ \citenamefont {Li}}]{Wang2024}%
	\BibitemOpen
	\bibfield  {author} {\bibinfo {author} {\bibfnamefont {J.}~\bibnamefont
			{Wang}}, \bibinfo {author} {\bibfnamefont {Y.}~\bibnamefont {Feng}}, \bibinfo
		{author} {\bibfnamefont {Y.}~\bibnamefont {Dai}}, \bibinfo {author}
		{\bibfnamefont {B.}~\bibnamefont {Huang}}, \bibinfo {author} {\bibfnamefont
			{Y.}~\bibnamefont {Ma}},\ and\ \bibinfo {author} {\bibfnamefont
			{X.}~\bibnamefont {Li}},\ }\bibfield  {title} {\bibinfo {title} {Pt symmetry
			breaking induced anomalous valley hall effect in 2d antiferromagnetic
			semiconductor},\ }\href {https://doi.org/10.1021/acs.jpclett.4c02451}
	{\bibfield  {journal} {\bibinfo  {journal} {The Journal of Physical Chemistry
				Letters}\ }\textbf {\bibinfo {volume} {15}},\ \bibinfo {pages} {9968}
		(\bibinfo {year} {2024})}\BibitemShut {NoStop}%
	\bibitem [{\citenamefont {Guo}\ \emph {et~al.}(2024)\citenamefont {Guo},
		\citenamefont {Tao}, \citenamefont {Zhuo}, \citenamefont {Zhu},\ and\
		\citenamefont {Ang}}]{Guo2024}%
	\BibitemOpen
	\bibfield  {author} {\bibinfo {author} {\bibfnamefont {S.-D.}\ \bibnamefont
			{Guo}}, \bibinfo {author} {\bibfnamefont {Y.-L.}\ \bibnamefont {Tao}},
		\bibinfo {author} {\bibfnamefont {Z.-Y.}\ \bibnamefont {Zhuo}}, \bibinfo
		{author} {\bibfnamefont {G.}~\bibnamefont {Zhu}},\ and\ \bibinfo {author}
		{\bibfnamefont {Y.~S.}\ \bibnamefont {Ang}},\ }\bibfield  {title} {\bibinfo
		{title} {Electric-field-tuned anomalous valley hall effect in a-type
			hexagonal antiferromagnetic monolayers},\ }\href
	{https://doi.org/10.1103/PhysRevB.109.134402} {\bibfield  {journal} {\bibinfo
			{journal} {Physical Review B}\ }\textbf {\bibinfo {volume} {109}},\ \bibinfo
		{pages} {134402} (\bibinfo {year} {2024})}\BibitemShut {NoStop}%
	\bibitem [{\citenamefont {Dong}\ \emph {et~al.}(2023)\citenamefont {Dong},
		\citenamefont {Jia}, \citenamefont {xiao Ji}, \citenamefont {shi Li},\ and\
		\citenamefont {Zhang}}]{Dong2023}%
	\BibitemOpen
	\bibfield  {author} {\bibinfo {author} {\bibfnamefont {X.-J.}\ \bibnamefont
			{Dong}}, \bibinfo {author} {\bibfnamefont {K.}~\bibnamefont {Jia}}, \bibinfo
		{author} {\bibfnamefont {W.}~\bibnamefont {xiao Ji}}, \bibinfo {author}
		{\bibfnamefont {S.}~\bibnamefont {shi Li}},\ and\ \bibinfo {author}
		{\bibfnamefont {C.-W.}\ \bibnamefont {Zhang}},\ }\bibfield  {title} {\bibinfo
		{title} {Valleytronics candidate with spontaneous valley polarization in
			a-type antiferromagnetic $mosi_2n_4$/$mnps_3$ heterostructure},\ }\href
	{https://doi.org/10.1021/acsaelm.2c01681} {\bibfield  {journal} {\bibinfo
			{journal} {ACS Applied Electronic Materials}\ }\textbf {\bibinfo {volume}
			{5}},\ \bibinfo {pages} {2046} (\bibinfo {year} {2023})}\BibitemShut
	{NoStop}%
	\bibitem [{\citenamefont {Xiao}\ \emph {et~al.}(2006)\citenamefont {Xiao},
		\citenamefont {Yao}, \citenamefont {Fang},\ and\ \citenamefont
		{Niu}}]{Xiao2006}%
	\BibitemOpen
	\bibfield  {author} {\bibinfo {author} {\bibfnamefont {D.}~\bibnamefont
			{Xiao}}, \bibinfo {author} {\bibfnamefont {Y.}~\bibnamefont {Yao}}, \bibinfo
		{author} {\bibfnamefont {Z.}~\bibnamefont {Fang}},\ and\ \bibinfo {author}
		{\bibfnamefont {Q.}~\bibnamefont {Niu}},\ }\bibfield  {title} {\bibinfo
		{title} {Berry-phase effect in anomalous thermoelectric transport},\ }\href
	{https://doi.org/10.1103/PhysRevLett.97.026603} {\bibfield  {journal}
		{\bibinfo  {journal} {Physical Review Letters}\ }\textbf {\bibinfo {volume}
			{97}},\ \bibinfo {pages} {26603} (\bibinfo {year} {2006})}\BibitemShut
	{NoStop}%
	\bibitem [{\citenamefont {Xiao}\ \emph {et~al.}(2012)\citenamefont {Xiao},
		\citenamefont {Liu}, \citenamefont {Feng}, \citenamefont {Xu},\ and\
		\citenamefont {Yao}}]{Xiao2012}%
	\BibitemOpen
	\bibfield  {author} {\bibinfo {author} {\bibfnamefont {D.}~\bibnamefont
			{Xiao}}, \bibinfo {author} {\bibfnamefont {G.-B.}\ \bibnamefont {Liu}},
		\bibinfo {author} {\bibfnamefont {W.}~\bibnamefont {Feng}}, \bibinfo {author}
		{\bibfnamefont {X.}~\bibnamefont {Xu}},\ and\ \bibinfo {author}
		{\bibfnamefont {W.}~\bibnamefont {Yao}},\ }\bibfield  {title} {\bibinfo
		{title} {Coupled spin and valley physics in monolayers of $mos_2$ and other
			group vi dichalcogenides},\ }\href
	{https://doi.org/10.1103/PhysRevLett.108.196802} {\bibfield  {journal}
		{\bibinfo  {journal} {Physical Review Letters}\ }\textbf {\bibinfo {volume}
			{108}},\ \bibinfo {pages} {196802} (\bibinfo {year} {2012})}\BibitemShut
	{NoStop}%
	\bibitem [{\citenamefont {Yamamoto}\ \emph {et~al.}(2015)\citenamefont
		{Yamamoto}, \citenamefont {Shimazaki}, \citenamefont {Borzenets},\ and\
		\citenamefont {Tarucha}}]{Yamamoto2015}%
	\BibitemOpen
	\bibfield  {author} {\bibinfo {author} {\bibfnamefont {M.}~\bibnamefont
			{Yamamoto}}, \bibinfo {author} {\bibfnamefont {Y.}~\bibnamefont {Shimazaki}},
		\bibinfo {author} {\bibfnamefont {I.~V.}\ \bibnamefont {Borzenets}},\ and\
		\bibinfo {author} {\bibfnamefont {S.}~\bibnamefont {Tarucha}},\ }\bibfield
	{title} {\bibinfo {title} {Valley hall effect in two dimensional hexagonal
			lattices},\ }\href {https://doi.org/10.7566/JPSJ.84.121006} {\bibfield
		{journal} {\bibinfo  {journal} {Journal of the Physical Society of Japan}\
		}\textbf {\bibinfo {volume} {84}},\ \bibinfo {pages} {121006} (\bibinfo
		{year} {2015})}\BibitemShut {NoStop}%
	\bibitem [{\citenamefont {Yu}\ \emph {et~al.}(2015)\citenamefont {Yu},
		\citenamefont {Zhu}, \citenamefont {Su},\ and\ \citenamefont
		{Jauho}}]{Yu2015}%
	\BibitemOpen
	\bibfield  {author} {\bibinfo {author} {\bibfnamefont {X.-Q.}\ \bibnamefont
			{Yu}}, \bibinfo {author} {\bibfnamefont {Z.-G.}\ \bibnamefont {Zhu}},
		\bibinfo {author} {\bibfnamefont {G.}~\bibnamefont {Su}},\ and\ \bibinfo
		{author} {\bibfnamefont {A.-P.}\ \bibnamefont {Jauho}},\ }\bibfield  {title}
	{\bibinfo {title} {Thermally driven pure spin and valley currents via the
			anomalous nernst effect in monolayer group-vi dichalcogenides},\ }\href
	{https://doi.org/10.1103/PhysRevLett.115.246601} {\bibfield  {journal}
		{\bibinfo  {journal} {Physical Review Letters}\ }\textbf {\bibinfo {volume}
			{115}},\ \bibinfo {pages} {246601} (\bibinfo {year} {2015})}\BibitemShut
	{NoStop}%
	\bibitem [{\citenamefont {Dau}\ \emph {et~al.}(2019)\citenamefont {Dau},
		\citenamefont {Vergnaud}, \citenamefont {Marty}, \citenamefont {Beigné},
		\citenamefont {Gambarelli}, \citenamefont {Maurel}, \citenamefont {Journot},
		\citenamefont {Hyot}, \citenamefont {Guillet}, \citenamefont {Grévin},
		\citenamefont {Okuno},\ and\ \citenamefont {Jamet}}]{Dau2019}%
	\BibitemOpen
	\bibfield  {author} {\bibinfo {author} {\bibfnamefont {M.~T.}\ \bibnamefont
			{Dau}}, \bibinfo {author} {\bibfnamefont {C.}~\bibnamefont {Vergnaud}},
		\bibinfo {author} {\bibfnamefont {A.}~\bibnamefont {Marty}}, \bibinfo
		{author} {\bibfnamefont {C.}~\bibnamefont {Beigné}}, \bibinfo {author}
		{\bibfnamefont {S.}~\bibnamefont {Gambarelli}}, \bibinfo {author}
		{\bibfnamefont {V.}~\bibnamefont {Maurel}}, \bibinfo {author} {\bibfnamefont
			{T.}~\bibnamefont {Journot}}, \bibinfo {author} {\bibfnamefont
			{B.}~\bibnamefont {Hyot}}, \bibinfo {author} {\bibfnamefont {T.}~\bibnamefont
			{Guillet}}, \bibinfo {author} {\bibfnamefont {B.}~\bibnamefont {Grévin}},
		\bibinfo {author} {\bibfnamefont {H.}~\bibnamefont {Okuno}},\ and\ \bibinfo
		{author} {\bibfnamefont {M.}~\bibnamefont {Jamet}},\ }\bibfield  {title}
	{\bibinfo {title} {The valley nernst effect in $wse_2$},\ }\href
	{https://doi.org/10.1038/s41467-019-13590-8} {\bibfield  {journal} {\bibinfo
			{journal} {Nature Communications}\ }\textbf {\bibinfo {volume} {10}},\
		\bibinfo {pages} {5796} (\bibinfo {year} {2019})}\BibitemShut {NoStop}%
	\bibitem [{\citenamefont {Du}\ \emph {et~al.}(2022)\citenamefont {Du},
		\citenamefont {Peng}, \citenamefont {He}, \citenamefont {Dai}, \citenamefont
		{Huang},\ and\ \citenamefont {Ma}}]{Du2022}%
	\BibitemOpen
	\bibfield  {author} {\bibinfo {author} {\bibfnamefont {W.}~\bibnamefont
			{Du}}, \bibinfo {author} {\bibfnamefont {R.}~\bibnamefont {Peng}}, \bibinfo
		{author} {\bibfnamefont {Z.}~\bibnamefont {He}}, \bibinfo {author}
		{\bibfnamefont {Y.}~\bibnamefont {Dai}}, \bibinfo {author} {\bibfnamefont
			{B.}~\bibnamefont {Huang}},\ and\ \bibinfo {author} {\bibfnamefont
			{Y.}~\bibnamefont {Ma}},\ }\bibfield  {title} {\bibinfo {title} {Anomalous
			valley hall effect in antiferromagnetic monolayers},\ }\href
	{https://doi.org/10.1038/s41699-022-00289-6} {\bibfield  {journal} {\bibinfo
			{journal} {npj 2D Materials and Applications}\ }\textbf {\bibinfo {volume}
			{6}},\ \bibinfo {pages} {11} (\bibinfo {year} {2022})}\BibitemShut {NoStop}%
	\bibitem [{\citenamefont {Bhowal}\ and\ \citenamefont
		{Vignale}(2021)}]{Bhowal2021}%
	\BibitemOpen
	\bibfield  {author} {\bibinfo {author} {\bibfnamefont {S.}~\bibnamefont
			{Bhowal}}\ and\ \bibinfo {author} {\bibfnamefont {G.}~\bibnamefont
			{Vignale}},\ }\bibfield  {title} {\bibinfo {title} {Orbital hall effect as an
			alternative to valley hall effect in gapped graphene},\ }\href
	{https://doi.org/10.1103/PhysRevB.103.195309} {\bibfield  {journal} {\bibinfo
			{journal} {Physical Review B}\ }\textbf {\bibinfo {volume} {103}},\ \bibinfo
		{pages} {195309} (\bibinfo {year} {2021})}\BibitemShut {NoStop}%
	\bibitem [{\citenamefont {Cysne}\ \emph {et~al.}(2021)\citenamefont {Cysne},
		\citenamefont {Costa}, \citenamefont {Canonico}, \citenamefont {Nardelli},
		\citenamefont {Muniz},\ and\ \citenamefont {Rappoport}}]{Cysne2021}%
	\BibitemOpen
	\bibfield  {author} {\bibinfo {author} {\bibfnamefont {T.~P.}\ \bibnamefont
			{Cysne}}, \bibinfo {author} {\bibfnamefont {M.}~\bibnamefont {Costa}},
		\bibinfo {author} {\bibfnamefont {L.~M.}\ \bibnamefont {Canonico}}, \bibinfo
		{author} {\bibfnamefont {M.~B.}\ \bibnamefont {Nardelli}}, \bibinfo {author}
		{\bibfnamefont {R.}~\bibnamefont {Muniz}},\ and\ \bibinfo {author}
		{\bibfnamefont {T.~G.}\ \bibnamefont {Rappoport}},\ }\bibfield  {title}
	{\bibinfo {title} {Disentangling orbital and valley hall effects in bilayers
			of transition metal dichalcogenides},\ }\href
	{https://doi.org/10.1103/PhysRevLett.126.056601} {\bibfield  {journal}
		{\bibinfo  {journal} {Physical Review Letters}\ }\textbf {\bibinfo {volume}
			{126}},\ \bibinfo {pages} {056601} (\bibinfo {year} {2021})}\BibitemShut
	{NoStop}%
	\bibitem [{\citenamefont {Go}\ \emph {et~al.}(2018)\citenamefont {Go},
		\citenamefont {Jo}, \citenamefont {Kim},\ and\ \citenamefont {Lee}}]{Go2018}%
	\BibitemOpen
	\bibfield  {author} {\bibinfo {author} {\bibfnamefont {D.}~\bibnamefont
			{Go}}, \bibinfo {author} {\bibfnamefont {D.}~\bibnamefont {Jo}}, \bibinfo
		{author} {\bibfnamefont {C.}~\bibnamefont {Kim}},\ and\ \bibinfo {author}
		{\bibfnamefont {H.-W.}\ \bibnamefont {Lee}},\ }\bibfield  {title} {\bibinfo
		{title} {Intrinsic spin and orbital hall effects from orbital texture},\
	}\href {https://doi.org/10.1103/PhysRevLett.121.086602} {\bibfield  {journal}
		{\bibinfo  {journal} {Physical Review Letters}\ }\textbf {\bibinfo {volume}
			{121}},\ \bibinfo {pages} {86602} (\bibinfo {year} {2018})}\BibitemShut
	{NoStop}%
	\bibitem [{\citenamefont {Kontani}\ \emph {et~al.}(2009)\citenamefont
		{Kontani}, \citenamefont {Tanaka}, \citenamefont {Hirashima}, \citenamefont
		{Yamada},\ and\ \citenamefont {Inoue}}]{Kontani2009}%
	\BibitemOpen
	\bibfield  {author} {\bibinfo {author} {\bibfnamefont {H.}~\bibnamefont
			{Kontani}}, \bibinfo {author} {\bibfnamefont {T.}~\bibnamefont {Tanaka}},
		\bibinfo {author} {\bibfnamefont {D.~S.}\ \bibnamefont {Hirashima}}, \bibinfo
		{author} {\bibfnamefont {K.}~\bibnamefont {Yamada}},\ and\ \bibinfo {author}
		{\bibfnamefont {J.}~\bibnamefont {Inoue}},\ }\bibfield  {title} {\bibinfo
		{title} {Giant orbital hall effect in transition metals: Origin of large spin
			and anomalous hall effects},\ }\href
	{https://doi.org/10.1103/PhysRevLett.102.016601} {\bibfield  {journal}
		{\bibinfo  {journal} {Physical Review Letters}\ }\textbf {\bibinfo {volume}
			{102}},\ \bibinfo {pages} {16601} (\bibinfo {year} {2009})}\BibitemShut
	{NoStop}%
	\bibitem [{\citenamefont {Sharma}\ and\ \citenamefont
		{Sarkar}(2024{\natexlab{a}})}]{Sharma12024}%
	\BibitemOpen
	\bibfield  {author} {\bibinfo {author} {\bibfnamefont {S.}~\bibnamefont
			{Sharma}}\ and\ \bibinfo {author} {\bibfnamefont {A.~D.}\ \bibnamefont
			{Sarkar}},\ }\bibfield  {title} {\bibinfo {title} {Harnessing orbital and
			valley thermal transport in 2d materials: The significance of inversion
			symmetry},\ }\bibfield  {journal} {\bibinfo  {journal} {Journal of Physics:
			Condensed Matter}\ }\href {https://doi.org/10.1088/1361-648X/ad9f0a}
	{10.1088/1361-648X/ad9f0a} (\bibinfo {year} {2024}{\natexlab{a}})\BibitemShut
	{NoStop}%
	\bibitem [{\citenamefont {Das}\ \emph {et~al.}(2024)\citenamefont {Das},
		\citenamefont {Ghorai}, \citenamefont {Culcer},\ and\ \citenamefont
		{Agarwal}}]{Das2024}%
	\BibitemOpen
	\bibfield  {author} {\bibinfo {author} {\bibfnamefont {K.}~\bibnamefont
			{Das}}, \bibinfo {author} {\bibfnamefont {K.}~\bibnamefont {Ghorai}},
		\bibinfo {author} {\bibfnamefont {D.}~\bibnamefont {Culcer}},\ and\ \bibinfo
		{author} {\bibfnamefont {A.}~\bibnamefont {Agarwal}},\ }\bibfield  {title}
	{\bibinfo {title} {Nonlinear valley hall effect},\ }\href
	{https://doi.org/10.1103/PhysRevLett.132.096302} {\bibfield  {journal}
		{\bibinfo  {journal} {Physical Review Letters}\ }\textbf {\bibinfo {volume}
			{132}},\ \bibinfo {pages} {96302} (\bibinfo {year} {2024})}\BibitemShut
	{NoStop}%
	\bibitem [{\citenamefont {Gao}\ \emph {et~al.}(2014)\citenamefont {Gao},
		\citenamefont {Yang},\ and\ \citenamefont {Niu}}]{Gao2014}%
	\BibitemOpen
	\bibfield  {author} {\bibinfo {author} {\bibfnamefont {Y.}~\bibnamefont
			{Gao}}, \bibinfo {author} {\bibfnamefont {S.~A.}\ \bibnamefont {Yang}},\ and\
		\bibinfo {author} {\bibfnamefont {Q.}~\bibnamefont {Niu}},\ }\bibfield
	{title} {\bibinfo {title} {Field induced positional shift of bloch electrons
			and its dynamical implications},\ }\href
	{https://doi.org/10.1103/PhysRevLett.112.166601} {\bibfield  {journal}
		{\bibinfo  {journal} {Physical Review Letters}\ }\textbf {\bibinfo {volume}
			{112}},\ \bibinfo {pages} {166601} (\bibinfo {year} {2014})}\BibitemShut
	{NoStop}%
	\bibitem [{\citenamefont {Liu}\ \emph {et~al.}(2021)\citenamefont {Liu},
		\citenamefont {Zhao}, \citenamefont {Huang}, \citenamefont {Wu},
		\citenamefont {Sheng}, \citenamefont {Xiao},\ and\ \citenamefont
		{Yang}}]{Liu2021}%
	\BibitemOpen
	\bibfield  {author} {\bibinfo {author} {\bibfnamefont {H.}~\bibnamefont
			{Liu}}, \bibinfo {author} {\bibfnamefont {J.}~\bibnamefont {Zhao}}, \bibinfo
		{author} {\bibfnamefont {Y.-X.}\ \bibnamefont {Huang}}, \bibinfo {author}
		{\bibfnamefont {W.}~\bibnamefont {Wu}}, \bibinfo {author} {\bibfnamefont
			{X.-L.}\ \bibnamefont {Sheng}}, \bibinfo {author} {\bibfnamefont
			{C.}~\bibnamefont {Xiao}},\ and\ \bibinfo {author} {\bibfnamefont {S.~A.}\
			\bibnamefont {Yang}},\ }\bibfield  {title} {\bibinfo {title} {Intrinsic
			second-order anomalous hall effect and its application in compensated
			antiferromagnets},\ }\href {https://doi.org/10.1103/PhysRevLett.127.277202}
	{\bibfield  {journal} {\bibinfo  {journal} {Physical Review Letters}\
		}\textbf {\bibinfo {volume} {127}},\ \bibinfo {pages} {277202} (\bibinfo
		{year} {2021})}\BibitemShut {NoStop}%
	\bibitem [{\citenamefont {Xiao}\ \emph
		{et~al.}(2021{\natexlab{a}})\citenamefont {Xiao}, \citenamefont {Liu},
		\citenamefont {Zhao}, \citenamefont {Yang},\ and\ \citenamefont
		{Niu}}]{Xiao2021}%
	\BibitemOpen
	\bibfield  {author} {\bibinfo {author} {\bibfnamefont {C.}~\bibnamefont
			{Xiao}}, \bibinfo {author} {\bibfnamefont {H.}~\bibnamefont {Liu}}, \bibinfo
		{author} {\bibfnamefont {J.}~\bibnamefont {Zhao}}, \bibinfo {author}
		{\bibfnamefont {S.~A.}\ \bibnamefont {Yang}},\ and\ \bibinfo {author}
		{\bibfnamefont {Q.}~\bibnamefont {Niu}},\ }\bibfield  {title} {\bibinfo
		{title} {Thermoelectric generation of orbital magnetization in metals},\
	}\href {https://doi.org/10.1103/PhysRevB.103.045401} {\bibfield  {journal}
		{\bibinfo  {journal} {Physical Review B}\ }\textbf {\bibinfo {volume}
			{103}},\ \bibinfo {pages} {45401} (\bibinfo {year}
		{2021}{\natexlab{a}})}\BibitemShut {NoStop}%
	\bibitem [{\citenamefont {Xiao}\ \emph
		{et~al.}(2021{\natexlab{b}})\citenamefont {Xiao}, \citenamefont {Ren},\ and\
		\citenamefont {Xiong}}]{Xiao12021}%
	\BibitemOpen
	\bibfield  {author} {\bibinfo {author} {\bibfnamefont {C.}~\bibnamefont
			{Xiao}}, \bibinfo {author} {\bibfnamefont {Y.}~\bibnamefont {Ren}},\ and\
		\bibinfo {author} {\bibfnamefont {B.}~\bibnamefont {Xiong}},\ }\bibfield
	{title} {\bibinfo {title} {Adiabatically induced orbital magnetization},\
	}\href {https://doi.org/10.1103/PhysRevB.103.115432} {\bibfield  {journal}
		{\bibinfo  {journal} {Physical Review B}\ }\textbf {\bibinfo {volume}
			{103}},\ \bibinfo {pages} {115432} (\bibinfo {year}
		{2021}{\natexlab{b}})}\BibitemShut {NoStop}%
	\bibitem [{\citenamefont {Tatara}(2015)}]{Tatara2015}%
	\BibitemOpen
	\bibfield  {author} {\bibinfo {author} {\bibfnamefont {G.}~\bibnamefont
			{Tatara}},\ }\bibfield  {title} {\bibinfo {title} {Thermal vector potential
			theory of transport induced by a temperature gradient},\ }\href
	{https://doi.org/10.1103/PhysRevLett.114.196601} {\bibfield  {journal}
		{\bibinfo  {journal} {Physical Review Letters}\ }\textbf {\bibinfo {volume}
			{114}},\ \bibinfo {pages} {196601} (\bibinfo {year} {2015})}\BibitemShut
	{NoStop}%
	\bibitem [{\citenamefont {Xiao}\ \emph {et~al.}(2010)\citenamefont {Xiao},
		\citenamefont {Chang},\ and\ \citenamefont {Niu}}]{Xiao2010}%
	\BibitemOpen
	\bibfield  {author} {\bibinfo {author} {\bibfnamefont {D.}~\bibnamefont
			{Xiao}}, \bibinfo {author} {\bibfnamefont {M.-C.}\ \bibnamefont {Chang}},\
		and\ \bibinfo {author} {\bibfnamefont {Q.}~\bibnamefont {Niu}},\ }\bibfield
	{title} {\bibinfo {title} {Berry phase effects on electronic properties},\
	}\href {https://doi.org/10.1103/RevModPhys.82.1959} {\bibfield  {journal}
		{\bibinfo  {journal} {Reviews of Modern Physics}\ }\textbf {\bibinfo {volume}
			{82}},\ \bibinfo {pages} {1959} (\bibinfo {year} {2010})}\BibitemShut
	{NoStop}%
	\bibitem [{\citenamefont {Sekine}\ \emph {et~al.}(2017)\citenamefont {Sekine},
		\citenamefont {Culcer},\ and\ \citenamefont {MacDonald}}]{Sekine2017}%
	\BibitemOpen
	\bibfield  {author} {\bibinfo {author} {\bibfnamefont {A.}~\bibnamefont
			{Sekine}}, \bibinfo {author} {\bibfnamefont {D.}~\bibnamefont {Culcer}},\
		and\ \bibinfo {author} {\bibfnamefont {A.~H.}\ \bibnamefont {MacDonald}},\
	}\bibfield  {title} {\bibinfo {title} {Quantum kinetic theory of the chiral
			anomaly},\ }\href {https://doi.org/10.1103/PhysRevB.96.235134} {\bibfield
		{journal} {\bibinfo  {journal} {Physical Review B}\ }\textbf {\bibinfo
			{volume} {96}},\ \bibinfo {pages} {235134} (\bibinfo {year}
		{2017})}\BibitemShut {NoStop}%
	\bibitem [{\citenamefont {Culcer}\ \emph {et~al.}(2012)\citenamefont {Culcer},
		\citenamefont {Saraiva}, \citenamefont {Koiller}, \citenamefont {Hu},\ and\
		\citenamefont {Sarma}}]{Culcer2012}%
	\BibitemOpen
	\bibfield  {author} {\bibinfo {author} {\bibfnamefont {D.}~\bibnamefont
			{Culcer}}, \bibinfo {author} {\bibfnamefont {A.~L.}\ \bibnamefont {Saraiva}},
		\bibinfo {author} {\bibfnamefont {B.}~\bibnamefont {Koiller}}, \bibinfo
		{author} {\bibfnamefont {X.}~\bibnamefont {Hu}},\ and\ \bibinfo {author}
		{\bibfnamefont {S.~D.}\ \bibnamefont {Sarma}},\ }\bibfield  {title} {\bibinfo
		{title} {Valley-based noise-resistant quantum computation using si quantum
			dots},\ }\href {https://doi.org/10.1103/PhysRevLett.108.126804} {\bibfield
		{journal} {\bibinfo  {journal} {Physical Review Letters}\ }\textbf {\bibinfo
			{volume} {108}},\ \bibinfo {pages} {126804} (\bibinfo {year}
		{2012})}\BibitemShut {NoStop}%
	\bibitem [{\citenamefont {Culcer}\ \emph {et~al.}(2017)\citenamefont {Culcer},
		\citenamefont {Sekine},\ and\ \citenamefont {MacDonald}}]{Culcer2017}%
	\BibitemOpen
	\bibfield  {author} {\bibinfo {author} {\bibfnamefont {D.}~\bibnamefont
			{Culcer}}, \bibinfo {author} {\bibfnamefont {A.}~\bibnamefont {Sekine}},\
		and\ \bibinfo {author} {\bibfnamefont {A.~H.}\ \bibnamefont {MacDonald}},\
	}\bibfield  {title} {\bibinfo {title} {Interband coherence response to
			electric fields in crystals: Berry-phase contributions and disorder
			effects},\ }\href {https://doi.org/10.1103/PhysRevB.96.035106} {\bibfield
		{journal} {\bibinfo  {journal} {Physical Review B}\ }\textbf {\bibinfo
			{volume} {96}},\ \bibinfo {pages} {35106} (\bibinfo {year}
		{2017})}\BibitemShut {NoStop}%
	\bibitem [{\citenamefont {Glazov}\ and\ \citenamefont
		{Ganichev}(2014)}]{Glazov2014}%
	\BibitemOpen
	\bibfield  {author} {\bibinfo {author} {\bibfnamefont {M.~M.}\ \bibnamefont
			{Glazov}}\ and\ \bibinfo {author} {\bibfnamefont {S.~D.}\ \bibnamefont
			{Ganichev}},\ }\bibfield  {title} {\bibinfo {title} {High frequency electric
			field induced nonlinear effects in graphene},\ }\href
	{https://doi.org/10.1016/j.physrep.2013.10.003} {\bibfield  {journal}
		{\bibinfo  {journal} {Physics Reports}\ }\textbf {\bibinfo {volume} {535}},\
		\bibinfo {pages} {101} (\bibinfo {year} {2014})}\BibitemShut {NoStop}%
	\bibitem [{\citenamefont {Watanabe}\ and\ \citenamefont
		{Yanase}(2021)}]{Watanabe2021}%
	\BibitemOpen
	\bibfield  {author} {\bibinfo {author} {\bibfnamefont {H.}~\bibnamefont
			{Watanabe}}\ and\ \bibinfo {author} {\bibfnamefont {Y.}~\bibnamefont
			{Yanase}},\ }\bibfield  {title} {\bibinfo {title} {Chiral photocurrent in
			parity-violating magnet and enhanced response in topological
			antiferromagnet},\ }\href {https://doi.org/10.1103/PhysRevX.11.011001}
	{\bibfield  {journal} {\bibinfo  {journal} {Physical Review X}\ }\textbf
		{\bibinfo {volume} {11}},\ \bibinfo {pages} {11001} (\bibinfo {year}
		{2021})}\BibitemShut {NoStop}%
	\bibitem [{\citenamefont {Varshney}\ \emph
		{et~al.}(2023{\natexlab{a}})\citenamefont {Varshney}, \citenamefont {Das},
		\citenamefont {Bhalla},\ and\ \citenamefont {Agarwal}}]{Varshney2023}%
	\BibitemOpen
	\bibfield  {author} {\bibinfo {author} {\bibfnamefont {H.}~\bibnamefont
			{Varshney}}, \bibinfo {author} {\bibfnamefont {K.}~\bibnamefont {Das}},
		\bibinfo {author} {\bibfnamefont {P.}~\bibnamefont {Bhalla}},\ and\ \bibinfo
		{author} {\bibfnamefont {A.}~\bibnamefont {Agarwal}},\ }\bibfield  {title}
	{\bibinfo {title} {Quantum kinetic theory of nonlinear thermal current},\
	}\href {https://doi.org/10.1103/PhysRevB.107.235419} {\bibfield  {journal}
		{\bibinfo  {journal} {Physical Review B}\ }\textbf {\bibinfo {volume}
			{107}},\ \bibinfo {pages} {235419} (\bibinfo {year}
		{2023}{\natexlab{a}})}\BibitemShut {NoStop}%
	\bibitem [{\citenamefont {Varshney}\ \emph
		{et~al.}(2023{\natexlab{b}})\citenamefont {Varshney}, \citenamefont
		{Mukherjee}, \citenamefont {Kundu},\ and\ \citenamefont
		{Agarwal}}]{Varshney12023}%
	\BibitemOpen
	\bibfield  {author} {\bibinfo {author} {\bibfnamefont {H.}~\bibnamefont
			{Varshney}}, \bibinfo {author} {\bibfnamefont {R.}~\bibnamefont {Mukherjee}},
		\bibinfo {author} {\bibfnamefont {A.}~\bibnamefont {Kundu}},\ and\ \bibinfo
		{author} {\bibfnamefont {A.}~\bibnamefont {Agarwal}},\ }\bibfield  {title}
	{\bibinfo {title} {Intrinsic nonlinear thermal hall transport of magnons: A
			quantum kinetic theory approach},\ }\href
	{https://doi.org/10.1103/PhysRevB.108.165412} {\bibfield  {journal} {\bibinfo
			{journal} {Physical Review B}\ }\textbf {\bibinfo {volume} {108}},\ \bibinfo
		{pages} {165412} (\bibinfo {year} {2023}{\natexlab{b}})}\BibitemShut
	{NoStop}%
	\bibitem [{\citenamefont {Das}\ \emph {et~al.}(2023)\citenamefont {Das},
		\citenamefont {Lahiri}, \citenamefont {Atencia}, \citenamefont {Culcer},\
		and\ \citenamefont {Agarwal}}]{Das2023}%
	\BibitemOpen
	\bibfield  {author} {\bibinfo {author} {\bibfnamefont {K.}~\bibnamefont
			{Das}}, \bibinfo {author} {\bibfnamefont {S.}~\bibnamefont {Lahiri}},
		\bibinfo {author} {\bibfnamefont {R.~B.}\ \bibnamefont {Atencia}}, \bibinfo
		{author} {\bibfnamefont {D.}~\bibnamefont {Culcer}},\ and\ \bibinfo {author}
		{\bibfnamefont {A.}~\bibnamefont {Agarwal}},\ }\bibfield  {title} {\bibinfo
		{title} {Intrinsic nonlinear conductivities induced by the quantum metric},\
	}\href {https://doi.org/10.1103/PhysRevB.108.L201405} {\bibfield  {journal}
		{\bibinfo  {journal} {Physical Review B}\ }\textbf {\bibinfo {volume}
			{108}},\ \bibinfo {pages} {L201405} (\bibinfo {year} {2023})}\BibitemShut
	{NoStop}%
	\bibitem [{\citenamefont {Hirata}\ \emph {et~al.}(2016)\citenamefont {Hirata},
		\citenamefont {Ishikawa}, \citenamefont {Miyagawa}, \citenamefont {Tamura},
		\citenamefont {Berthier}, \citenamefont {Basko}, \citenamefont {Kobayashi},
		\citenamefont {Matsuno},\ and\ \citenamefont {Kanoda}}]{Hirata2016}%
	\BibitemOpen
	\bibfield  {author} {\bibinfo {author} {\bibfnamefont {M.}~\bibnamefont
			{Hirata}}, \bibinfo {author} {\bibfnamefont {K.}~\bibnamefont {Ishikawa}},
		\bibinfo {author} {\bibfnamefont {K.}~\bibnamefont {Miyagawa}}, \bibinfo
		{author} {\bibfnamefont {M.}~\bibnamefont {Tamura}}, \bibinfo {author}
		{\bibfnamefont {C.}~\bibnamefont {Berthier}}, \bibinfo {author}
		{\bibfnamefont {D.}~\bibnamefont {Basko}}, \bibinfo {author} {\bibfnamefont
			{A.}~\bibnamefont {Kobayashi}}, \bibinfo {author} {\bibfnamefont
			{G.}~\bibnamefont {Matsuno}},\ and\ \bibinfo {author} {\bibfnamefont
			{K.}~\bibnamefont {Kanoda}},\ }\bibfield  {title} {\bibinfo {title}
		{Observation of an anisotropic dirac cone reshaping and ferrimagnetic spin
			polarization in an organic conductor},\ }\href
	{https://doi.org/10.1038/ncomms12666} {\bibfield  {journal} {\bibinfo
			{journal} {Nature Communications}\ }\textbf {\bibinfo {volume} {7}},\
		\bibinfo {pages} {12666} (\bibinfo {year} {2016})}\BibitemShut {NoStop}%
	\bibitem [{\citenamefont {Mojarro}\ \emph {et~al.}(2021)\citenamefont
		{Mojarro}, \citenamefont {Carrillo-Bastos},\ and\ \citenamefont
		{Maytorena}}]{Mojarro2021}%
	\BibitemOpen
	\bibfield  {author} {\bibinfo {author} {\bibfnamefont {M.~A.}\ \bibnamefont
			{Mojarro}}, \bibinfo {author} {\bibfnamefont {R.}~\bibnamefont
			{Carrillo-Bastos}},\ and\ \bibinfo {author} {\bibfnamefont {J.~A.}\
			\bibnamefont {Maytorena}},\ }\bibfield  {title} {\bibinfo {title} {Optical
			properties of massive anisotropic tilted dirac systems},\ }\href
	{https://doi.org/10.1103/PhysRevB.103.165415} {\bibfield  {journal} {\bibinfo
			{journal} {Physical Review B}\ }\textbf {\bibinfo {volume} {103}},\ \bibinfo
		{pages} {165415} (\bibinfo {year} {2021})}\BibitemShut {NoStop}%
	\bibitem [{\citenamefont {Mojarro}\ \emph {et~al.}(2022)\citenamefont
		{Mojarro}, \citenamefont {Carrillo-Bastos},\ and\ \citenamefont
		{Maytorena}}]{Mojarro2022}%
	\BibitemOpen
	\bibfield  {author} {\bibinfo {author} {\bibfnamefont {M.~A.}\ \bibnamefont
			{Mojarro}}, \bibinfo {author} {\bibfnamefont {R.}~\bibnamefont
			{Carrillo-Bastos}},\ and\ \bibinfo {author} {\bibfnamefont {J.~A.}\
			\bibnamefont {Maytorena}},\ }\bibfield  {title} {\bibinfo {title} {Hyperbolic
			plasmons in massive tilted two-dimensional dirac materials},\ }\href
	{https://doi.org/10.1103/PhysRevB.105.L201408} {\bibfield  {journal}
		{\bibinfo  {journal} {Physical Review B}\ }\textbf {\bibinfo {volume}
			{105}},\ \bibinfo {pages} {L201408} (\bibinfo {year} {2022})}\BibitemShut
	{NoStop}%
	\bibitem [{\citenamefont {Jr}\ \emph {et~al.}(2009)\citenamefont {Jr},
		\citenamefont {Peeters}, \citenamefont {Filho},\ and\ \citenamefont
		{Farias}}]{Jr2009}%
	\BibitemOpen
	\bibfield  {author} {\bibinfo {author} {\bibfnamefont {J.~M.~P.}\
			\bibnamefont {Jr}}, \bibinfo {author} {\bibfnamefont {F.~M.}\ \bibnamefont
			{Peeters}}, \bibinfo {author} {\bibfnamefont {R.~N.~C.}\ \bibnamefont
			{Filho}},\ and\ \bibinfo {author} {\bibfnamefont {G.~A.}\ \bibnamefont
			{Farias}},\ }\bibfield  {title} {\bibinfo {title} {Valley polarization due to
			trigonal warping on tunneling electrons in graphene},\ }\href
	{https://doi.org/10.1088/0953-8984/21/4/045301} {\bibfield  {journal}
		{\bibinfo  {journal} {Journal of Physics: Condensed Matter}\ }\textbf
		{\bibinfo {volume} {21}},\ \bibinfo {pages} {45301} (\bibinfo {year}
		{2009})}\BibitemShut {NoStop}%
	\bibitem [{\citenamefont {Cserti}\ \emph {et~al.}(2007)\citenamefont {Cserti},
		\citenamefont {Csordás},\ and\ \citenamefont {Dávid}}]{Cserti2007}%
	\BibitemOpen
	\bibfield  {author} {\bibinfo {author} {\bibfnamefont {J.}~\bibnamefont
			{Cserti}}, \bibinfo {author} {\bibfnamefont {A.}~\bibnamefont {Csordás}},\
		and\ \bibinfo {author} {\bibfnamefont {G.}~\bibnamefont {Dávid}},\
	}\bibfield  {title} {\bibinfo {title} {Role of the trigonal warping on the
			minimal conductivity of bilayer graphene},\ }\href
	{https://doi.org/10.1103/PhysRevLett.99.066802} {\bibfield  {journal}
		{\bibinfo  {journal} {Physical Review Letters}\ }\textbf {\bibinfo {volume}
			{99}},\ \bibinfo {pages} {66802} (\bibinfo {year} {2007})}\BibitemShut
	{NoStop}%
	\bibitem [{\citenamefont {Kormányos}\ \emph {et~al.}(2013)\citenamefont
		{Kormányos}, \citenamefont {Zólyomi}, \citenamefont {Drummond},
		\citenamefont {Rakyta}, \citenamefont {Burkard},\ and\ \citenamefont
		{Fal'ko}}]{Andor2013}%
	\BibitemOpen
	\bibfield  {author} {\bibinfo {author} {\bibfnamefont {A.}~\bibnamefont
			{Kormányos}}, \bibinfo {author} {\bibfnamefont {V.}~\bibnamefont
			{Zólyomi}}, \bibinfo {author} {\bibfnamefont {N.~D.}\ \bibnamefont
			{Drummond}}, \bibinfo {author} {\bibfnamefont {P.}~\bibnamefont {Rakyta}},
		\bibinfo {author} {\bibfnamefont {G.}~\bibnamefont {Burkard}},\ and\ \bibinfo
		{author} {\bibfnamefont {V.~I.}\ \bibnamefont {Fal'ko}},\ }\bibfield  {title}
	{\bibinfo {title} {Monolayer $mos_2$ : Trigonal warping, the $\gamma$ valley,
			and spin-orbit coupling effects},\ }\href
	{https://doi.org/10.1103/PhysRevB.88.045416} {\bibfield  {journal} {\bibinfo
			{journal} {Physical Review B}\ }\textbf {\bibinfo {volume} {88}},\ \bibinfo
		{pages} {45416} (\bibinfo {year} {2013})}\BibitemShut {NoStop}%
	\bibitem [{\citenamefont {Wu}\ \emph {et~al.}(2021)\citenamefont {Wu},
		\citenamefont {Zhu},\ and\ \citenamefont {Yu}}]{Wu2021}%
	\BibitemOpen
	\bibfield  {author} {\bibinfo {author} {\bibfnamefont {Y.-L.}\ \bibnamefont
			{Wu}}, \bibinfo {author} {\bibfnamefont {G.-H.}\ \bibnamefont {Zhu}},\ and\
		\bibinfo {author} {\bibfnamefont {X.-Q.}\ \bibnamefont {Yu}},\ }\bibfield
	{title} {\bibinfo {title} {Nonlinear anomalous nernst effect in strained
			graphene induced by trigonal warping},\ }\href
	{https://doi.org/10.1103/PhysRevB.104.195427} {\bibfield  {journal} {\bibinfo
			{journal} {Physical Review B}\ }\textbf {\bibinfo {volume} {104}},\ \bibinfo
		{pages} {195427} (\bibinfo {year} {2021})}\BibitemShut {NoStop}%
	\bibitem [{\citenamefont {Winkler}(2003)}]{Winkler2003}%
	\BibitemOpen
	\bibfield  {author} {\bibinfo {author} {\bibfnamefont {R.}~\bibnamefont
			{Winkler}},\ }\href {https://doi.org/10.1007/b13586} {\emph {\bibinfo {title}
			{Spin—Orbit Coupling Effects in Two-Dimensional Electron and Hole
				Systems}}},\ Vol.\ \bibinfo {volume} {191}\ (\bibinfo  {publisher} {Springer
		Berlin Heidelberg},\ \bibinfo {year} {2003})\BibitemShut {NoStop}%
	\bibitem [{\citenamefont {Kechedzhi}\ \emph {et~al.}(2007)\citenamefont
		{Kechedzhi}, \citenamefont {Fal’ko}, \citenamefont {McCann},\ and\
		\citenamefont {Altshuler}}]{Kechedzhi2007}%
	\BibitemOpen
	\bibfield  {author} {\bibinfo {author} {\bibfnamefont {K.}~\bibnamefont
			{Kechedzhi}}, \bibinfo {author} {\bibfnamefont {V.~I.}\ \bibnamefont
			{Fal’ko}}, \bibinfo {author} {\bibfnamefont {E.}~\bibnamefont {McCann}},\
		and\ \bibinfo {author} {\bibfnamefont {B.~L.}\ \bibnamefont {Altshuler}},\
	}\bibfield  {title} {\bibinfo {title} {Influence of trigonal warping on
			interference effects in bilayer graphene},\ }\href
	{https://doi.org/10.1103/PhysRevLett.98.176806} {\bibfield  {journal}
		{\bibinfo  {journal} {Physical Review Letters}\ }\textbf {\bibinfo {volume}
			{98}},\ \bibinfo {pages} {176806} (\bibinfo {year} {2007})}\BibitemShut
	{NoStop}%
	\bibitem [{\citenamefont {Goerbig}\ \emph {et~al.}(2008)\citenamefont
		{Goerbig}, \citenamefont {Fuchs}, \citenamefont {Montambaux},\ and\
		\citenamefont {Piechon}}]{Goerbig2008}%
	\BibitemOpen
	\bibfield  {author} {\bibinfo {author} {\bibfnamefont {M.}~\bibnamefont
			{Goerbig}}, \bibinfo {author} {\bibfnamefont {J.-N.}\ \bibnamefont {Fuchs}},
		\bibinfo {author} {\bibfnamefont {G.}~\bibnamefont {Montambaux}},\ and\
		\bibinfo {author} {\bibfnamefont {F.}~\bibnamefont {Piechon}},\ }\bibfield
	{title} {\bibinfo {title} {Tilted anisotropic dirac cones in quinoid-type
			graphene and $\alpha({B}{E}{D}{T}-{T}{T}{F})_2{I}_3$},\ }\href
	{https://doi.org/10.1103/PhysRevB.78.045415} {\bibfield  {journal} {\bibinfo
			{journal} {Physical Review B}\ }\textbf {\bibinfo {volume} {78}},\ \bibinfo
		{pages} {045415} (\bibinfo {year} {2008})}\BibitemShut {NoStop}%
	\bibitem [{\citenamefont {Kobayashi}\ \emph {et~al.}(2007)\citenamefont
		{Kobayashi}, \citenamefont {Katayama}, \citenamefont {Suzumura},\ and\
		\citenamefont {Fukuyama}}]{Kobayashi2007}%
	\BibitemOpen
	\bibfield  {author} {\bibinfo {author} {\bibfnamefont {A.}~\bibnamefont
			{Kobayashi}}, \bibinfo {author} {\bibfnamefont {S.}~\bibnamefont {Katayama}},
		\bibinfo {author} {\bibfnamefont {Y.}~\bibnamefont {Suzumura}},\ and\
		\bibinfo {author} {\bibfnamefont {H.}~\bibnamefont {Fukuyama}},\ }\bibfield
	{title} {\bibinfo {title} {Massless fermions in organic conductor},\ }\href
	{https://doi.org/10.1143/JPSJ.76.034711} {\bibfield  {journal} {\bibinfo
			{journal} {Journal of the Physical Society of Japan}\ }\textbf {\bibinfo
			{volume} {76}},\ \bibinfo {pages} {034711} (\bibinfo {year}
		{2007})}\BibitemShut {NoStop}%
	\bibitem [{\citenamefont {Xiao}\ \emph {et~al.}(2007)\citenamefont {Xiao},
		\citenamefont {Yao},\ and\ \citenamefont {Niu}}]{Xiao2007}%
	\BibitemOpen
	\bibfield  {author} {\bibinfo {author} {\bibfnamefont {D.}~\bibnamefont
			{Xiao}}, \bibinfo {author} {\bibfnamefont {W.}~\bibnamefont {Yao}},\ and\
		\bibinfo {author} {\bibfnamefont {Q.}~\bibnamefont {Niu}},\ }\bibfield
	{title} {\bibinfo {title} {Valley-contrasting physics in graphene: Magnetic
			moment and topological transport},\ }\href
	{https://doi.org/10.1103/PhysRevLett.99.236809} {\bibfield  {journal}
		{\bibinfo  {journal} {Physical Review Letters}\ }\textbf {\bibinfo {volume}
			{99}},\ \bibinfo {pages} {236809} (\bibinfo {year} {2007})}\BibitemShut
	{NoStop}%
	\bibitem [{\citenamefont {Neumann}\ \emph {et~al.}(2022)\citenamefont
		{Neumann}, \citenamefont {Mook}, \citenamefont {Henk},\ and\ \citenamefont
		{Mertig}}]{Neumann2022}%
	\BibitemOpen
	\bibfield  {author} {\bibinfo {author} {\bibfnamefont {R.~R.}\ \bibnamefont
			{Neumann}}, \bibinfo {author} {\bibfnamefont {A.}~\bibnamefont {Mook}},
		\bibinfo {author} {\bibfnamefont {J.}~\bibnamefont {Henk}},\ and\ \bibinfo
		{author} {\bibfnamefont {I.}~\bibnamefont {Mertig}},\ }\bibfield  {title}
	{\bibinfo {title} {Thermal hall effect of magnons in collinear
			antiferromagnetic insulators: Signatures of magnetic and topological phase
			transitions},\ }\href {https://doi.org/10.1103/PhysRevLett.128.117201}
	{\bibfield  {journal} {\bibinfo  {journal} {Physical Review Letters}\
		}\textbf {\bibinfo {volume} {128}},\ \bibinfo {pages} {117201} (\bibinfo
		{year} {2022})}\BibitemShut {NoStop}%
	\bibitem [{\citenamefont {Sharma}\ and\ \citenamefont
		{Sarkar}(2024{\natexlab{b}})}]{Sharma2024}%
	\BibitemOpen
	\bibfield  {author} {\bibinfo {author} {\bibfnamefont {S.}~\bibnamefont
			{Sharma}}\ and\ \bibinfo {author} {\bibfnamefont {A.~D.}\ \bibnamefont
			{Sarkar}},\ }\bibfield  {title} {\bibinfo {title} {Conflux of spin nernst and
			spin hall effect in the $zncu_2snse_4$ topological insulator},\ }\href
	{https://doi.org/10.1088/1361-648X/ad68b5} {\bibfield  {journal} {\bibinfo
			{journal} {J. Phys.: Condens. Matter}\ }\textbf {\bibinfo {volume} {36}},\
		\bibinfo {pages} {445501} (\bibinfo {year} {2024}{\natexlab{b}})}\BibitemShut
	{NoStop}%
\end{thebibliography}
\end{document}